\documentclass[sts]{imsart}

\RequirePackage{amsthm,amsmath,amsfonts,amssymb}
\RequirePackage[numbers,sort&compress]{natbib}
\RequirePackage[colorlinks,citecolor=blue,urlcolor=blue]{hyperref}
\RequirePackage{graphicx}
\usepackage{rotating}
\usepackage{tcolorbox}

\definecolor{explaincolor}{RGB}{42, 161, 152}      
\definecolor{modelcolor}{RGB}{38, 139, 210}     
\definecolor{analysiscolor}{RGB}{181, 73, 91}   
\definecolor{datacolor}{RGB}{147, 112, 219}

\startlocaldefs
\theoremstyle{plain}

\theoremstyle{remark}

\endlocaldefs

\begin{document}

\begin{frontmatter}
\title{Recent advances in doubly-robust weighted ordinary least squares techniques for dynamic treatment regime estimation}

\runtitle{Advances in dWOLS-based DTR estimation techniques}

\begin{aug}
\author[A]{\fnms{Adel}~\snm{Ahmadi Nadi}\thanks{[\textbf{Corresponding author indication should be put in the Acknowledgment section if necessary.}]}\ead[label=e1]{adel.ahmadinadi@uwaterloo.ca}}
\and
\author[B]{\fnms{Michael}~\snm{Wallace}\ead[label=e2]{michael.wallace@uwaterloo.ca}
}


\address[A]{Adel Ahmadi Nadi is a Postdoctoral fellow in the Department of Statistics and Actuarial Science at the University of Waterloo, in Waterloo, Ontario, Canada\printead[presep={\ }]{e1}.}

\address[B]{Michael Michael is an Associate Professor in the Department of Statistics and Actuarial Science at the University of Waterloo, in Waterloo, Ontario, Canada.\printead[presep={\ }]{e2}.}

\end{aug}

\begin{abstract}
A dynamic treatment regime (DTR) is an approach to delivering precision medicine that uses patient characteristics to guide treatment decisions for optimal health outcomes. Numerous methods have been proposed for DTR estimation, including dynamic weighted ordinary least squares (dWOLS), a regression-based approach that affords double robustness to model misspecification within an easy to implement analytical framework. Initially, the dWOLS approach was developed under the assumptions of continuous outcomes and binary treatment decisions. Motivated by clinical research, subsequent theoretical advancements have extended the dWOLS framework to address binary, continuous and multicategory treatments across various outcome types, including binary, continuous, and survival-type. However, certain scenarios remain unexplored. This paper summarizes the last ten years of extension and application of the dWOLS method, providing a comprehensive and detailed review of the original dWOLS method and its extensions, as well as highlighting its diverse practical applications. We also explore studies that have addressed challenges associated with dWOLS implementation, such as model validation, variable selection, and handling measurement errors. Using simulated data, we present numerical illustrations along with step-by-step implementations in the \texttt{R} environment to facilitate a deeper understanding of dWOLS-based DTR estimation methodologies.
\end{abstract}


\end{frontmatter}

\section{Introduction} 
Precision (also known as personalized) medicine enhances patient outcomes by tailoring treatments to individual characteristics and responses. Unlike standardized "one-size-fits-all" approaches, precision medicine uses evolving patient data to prescribe a personalized sequence of treatment rules for each individual.  This approach is particularly effective in managing chronic diseases within heterogeneous patient populations, where multiple treatment stages and follow-ups are required. The benefits of personalized medicine include improved treatment efficacy, fewer side effects, better patient adherence, and reduced healthcare costs \cite{chakraborty2013statistical}.

Dynamic treatment regimes (DTRs) are one approach to operationalizing precision medicine. DTRs are sequences of treatment decision rules, taking patient characteristics as inputs and outputting treatment recommendations. These characteristics may encompass the patient’s current health status (including biomarkers and characteristics), historical health information (such as responses to and side effects from previous treatments), and potential future outcomes (anticipated responses to future treatments). By leveraging this data-driven approach, DTRs aim to maximize health benefits for patients \cite{moodie2020precision}. As a systematic approach to individualized care, DTRs are gaining more attention in clinical research and are recognized for enhancing patient-centred healthcare \cite{chakraborty2014dynamic}.

There are two primary classes of statistical approaches for optimally estimating a DTR: value search (direct) methods and regression-based (indirect) approaches. Within the regression-based class, two well-established approaches are G-estimation \cite{robins2004optimal} and Q-learning \cite{moodie2014q}. Q-learning, grounded in regression modeling, offers implementational and conceptual simplicity, but little robustness to model misspecification. G-estimation, meanwhile, is based on estimating equations and affords greater robustness but more challenging implementation.

Wallace and Moodie \cite{wallace2015doubly} developed the dynamic weighted ordinary least squares (dWOLS) approach for DTR estimation which combines the ease of implementation characteristic of Q-learning with the robustness in modeling offered by G-estimation. Since its introduction, the dWOLS approach has attracted significant attention, leading to both theoretical advancements and practical applications. To further support practitioners, Wallace et al. \cite{wallace2016package,wallace2017dynamic} introduced the DTRreg \texttt{R} package, which facilitates the application of dWOLS to various problems, making the method more accessible to researchers and clinicians. Since then, dWOLS-based methods have been applied to numerous DTR estimation problems. 

The original dWOLS method was introduced with certain assumptions about the data, some of which may vary or be violated in different contexts. For example, it assumes that the outcome of interest measured on each individual is continuous (e.g., blood pressure). However, many clinical studies involve outcomes that are binary, categorical, or represent survival times with the possibility of censoring. Moreover, the dWOLS framework assumes that treatment decisions at each decision point are binary. In contrast, many clinical scenarios involve more complex decision-making processes, such as selecting from multiple treatment options at each decision point (categorical treatments) or determining an individualized drug dosage (continuous treatments). The dWOLS method also relies on some other assumptions, including a predefined functional relationship between the outcome and patient data, the absence of measurement error in observations, and the absence of interference (i.e., an individual’s outcome is not influenced by the treatments assigned to others). These assumptions may be violated in practice, posing challenges to the method's accuracy and applicability. 

Motivated by clinical studies, subsequent theoretical advancements have extended the dWOLS framework to accommodate binary, continuous and multicategory treatments across various outcome types, including binary, continuous, and survival-type. Additionally, studies have addressed challenges associated with dWOLS implementation, such as model validation, variable selection, and the handling of measurement errors and interference. 

With the growing emphasis on personalized medicine and dynamic treatment regimes, understanding methods like dWOLS has become increasingly important. This can be challenging in a large and varied literature, with methods and applications presented across a wide range of contexts and to a wide variety of audiences. This paper, coming ten years after the dWOLS method was first developed, provides a comprehensive review of dWOLS-based DTR estimation methods, balancing theoretical foundations with practical implications. While theoretical advancements are presented in a detailed yet accessible manner, methodological explanations are simplified to aid clarity. We also explore studies that have successfully applied dWOLS-based approaches to real-world problems, offering insights into their implementation. To further enhance understanding, we present examples using simulated data, along with step-by-step implementations of various dWOLS-based methods in the \texttt{R} environment. These efforts aim to make the methodologies more intuitive and easier to apply to specific problems, given the often complex theoretical aspects of the existing techniques. This paper therefore serves as a resource for (bio)statisticians, clinicians, and researchers, who are interested in the current, complete, state of the art of the dWOLS approach and its extensions and applications.

 The remainder of this paper is organized as follows. Section 2 reviews the original dWOLS approach, covering the notation, statistical models, and implementation procedures. Section 3 explores direct extensions to the dWOLS methodology. Section 4 examines studies addressing cases where the original dWOLS assumptions are violated or modified. Section 5 and Appendices A-D present numerical illustrations of the dWOLS-based approaches. Finally, Section 6 discusses research gaps and potential directions for future research.

\begin{table*}[]
 \small
 \setlength{\tabcolsep}{8pt}
\renewcommand{\arraystretch}{1}
\caption{Table of notations.}
\begin{tabular}{ll}
\hline\hline
Sizes and Indices & Description  \\
\hline
$n$ & The number of individuals participating in the study \\
$i$ &    Index for individuals: $i=1, 2,\ldots,n$  \\
$K$ & The number of treatment regime stages \\
$j$ &    Index for the stages of treatment regime: $j=1, 2,\ldots,K$  \\
$N_j$ & The number of available treatment options at stage $j$ \\
\hline
Random (observed) variables & Description  \\
\hline
$Y\, (y)$ & The random (observed) outcome of interest      \\
$Y^{\mathbf{a}} $ & The potential (counterfactual) outcome under the treatment regime $\mathbf{a}$   \\
$\tilde{Y}_j\, (\tilde{y}_j)$ & The (observed) pseudo outcome at stage $j$ \\
$C$ & Random censoring time \\
$\Delta\, (\delta)$ & Random (observed) censoring indicator with $\Delta=0$ meaning censored data \\
$\epsilon$ & Random error term in the accelerated failure time model \\
$U$ &  Error term in the measurement error context \\
$\eta_j$ & The indicator with $\eta_j=1$ meaning the individual entered stage $j$ of a treatment regime\\
\hline
Treatment & Description  \\
\hline
$a$ & Treatment decision taking value from the treatment space $\mathcal{A}$ \\
$a^{opt}$ & Optimal treatment decision \\
$\overline{\mathbf{a}}_j=(a_1,\ldots,a_j)$ & Treatment decisions (regime) up to stage $j$  \\
$\underline{\mathbf{a}}_{j}=(a_{j},\ldots,a_K)$ & Treatment decisions (regime) from stage $j$ to the final stage $K$ \\
\hline
Information & Description  \\
\hline
$\mathbf{x}\, (\mathbf{x}_j)$ & Covariate matrix (up to stage $j$) \\
$\mathbf{x}^*$ & Error-prone covariate matrix  \\
$\mathbf{h}\, (\mathbf{h}_j)$ & Full (up to stage $j$) information matrix  \\
$\mathbf{h}^{\pmb{\beta}}\, (\mathbf{h}_j^{\pmb{\beta}})$  & Full (up to stage $j$) non-treatment covariates matrix \\
$\mathbf{h}^{\pmb{\Psi}}\, (\mathbf{h}_j^{\pmb{\Psi}})$  & Full (up to stage $j$) tailoring covariates matrix \\
\hline
Statistical models  & Description \\
\hline
$\mathbb{E}(Y^{\overline{\mathbf{a}}_K}|\mathbf{h}; \pmb{\beta}, \pmb{\Psi})$ &  Outcom model (sometimes shown by $Q$)  \\
$f_j\left(\mathbf{h}_j^{\pmb{\beta}}; \pmb{\beta}_j\right)$ &  Treatment-free model at stage $j$ \\
$\gamma_j(\mathbf{h}_j^{\Psi}, a_j; \pmb{\Psi}_j)$ & Blip model at stage $j$ \\
$\mu_j(\mathbf{h}_j^{\Psi}, a_j; \pmb{\Psi}_j)$ & Regret model at stage $j$ \\
$\pi(\mathbf{h}; \pmb{\alpha})=\mathbb{P}(A=1|\mathbf{h};\pmb{\alpha})$ & Treatment model with binary treatment decisions, i.e., when $a \in \{0, 1\}$ \\
$\pi(\mathbf{h}, a; \pmb{\alpha})=\mathbb{P}(A=a|\mathbf{h}, \eta=1;\pmb{\alpha})$ & Treatment model with categorical treatment decisions, i.e., when $a \in \{1, 2, \ldots, N\}$ \\
$\pi(\mathbf{h}, a; \pmb{\alpha})=f_{A|\mathbf{h}}(a|\mathbf{h}; \pmb{\alpha})$ & Generalized propensity score model with continuous treatments \\
$g(\mathbf{h}, a)=\mathbb{P}(\Delta=0|\mathbf{h}, a, \eta=1)$ & Censoring model \\
$w(\mathbf{h}, a)$ or $w(\mathbf{h}, a, \delta)$ & Balancing weight \\
$m(a)$ & Treatment-specific weight \\
\hline\hline
\end{tabular}
\begin{flushleft}
\small
\textbf{Note:} Vectors and matrices are represented in bold font, uppercase letters indicate random variables, lowercase letters correspond to their realizations, and overline and underline notations show historical and future values, respectively.
\end{flushleft}
\label{tab:notations}
\end{table*}

\section{DTR estimation using the dWOLS approach}
\label{sec:dWOLS}
We begin with an overview of the dWOLS method, including establishing our notation and a detailed description of the estimation procedure and its interpretation.

\subsection*{Notations and blip model}

We consider treatment regimes with $K \geq 1$ stages, where each stage begins with a treatment decision.  Vectors and matrices are presented in bold font throughout the paper. Unless otherwise specified, uppercase letters denote random variables, while lowercase letters represent the realizations of these random variables. We also define
\begin{itemize}
    \item $y$: The continuous outcome of interest which is typically defined as a higher-the-better outcome and can be observed (measured) at the end of the last stage of the regime (stage $K$).
    \item $a_j$: The  binary treatment decision at stage $j$, for $j=1, 2, \ldots, K$, that takes values in $\mathcal{A}_j=\{0, 1\}$. In this setting, $a_j = 1$ indicates a treatment or intervention, and $a_j = 0$ indicates no treatment, a control, or standard care. 
  \item $\mathbf{x}_j$: Non-treatment information (covariates) available before the $j$-th treatment decision (e.g., age or disease severity).
    \item $\mathbf{h}_j$: The patient's history up to stage $j$, including past treatments and covariates. This contains all the information available prior to the stage $j$ treatment decision being made.
\end{itemize}

When appropriate, the subscript $j$ stage notation is suppressed. To be consistent with the existing literature, especially Wallace and Moodie \cite{wallace2015doubly}, we use overline notation to denote history and underline notation for future values. For example,  $\overline{\mathbf{a}}_j=(a_1,\ldots,a_j)$  represents all treatment indicators up to and including stage $j$, while $\underline{\mathbf{a}}_{j+1}=(a_{j+1},\ldots,a_K)$ captures the treatment decisions from stage $j+1$ to the final stage $K$. The matrix $\mathbf{h}_j$ can thus be represented as $\mathbf{h}_j = (\overline{\mathbf{x}}_j, \overline{\mathbf{a}}_{j-1})$, where $\overline{\mathbf{x}}_j$ refers to the covariates observed up to and including stage $j$. We further introduce the notation $Y^{a}$, known as the counterfactual or potential outcome, representing the outcome for a patient who receives treatment $a$. This notation denotes the hypothetical outcome that would occur if, contrary to fact, the patient followed the specified treatment regimen, even though an identical patient with the same history and treatment may not exist in the population.

dWOLS relies on some assumptions well-known in causal inference: the stable unit treatment value assumption (SUTVA) and the assumptions of no unmeasured confounders, positivity, and consistency.
\begin{itemize}
    \item \( A.1 \): SUTVA implies that the outcome for any individual is not influenced by the treatment assignments of other individuals, i.e., the absence of interference. In other words, SUTVA assumes that each individual's outcome depends solely on their own covariates and the treatments they receive.
    \item \( A.2 \): The assumption of no unmeasured confounders asserts that the treatment assignment at any stage $j$ is independent of future covariates or outcomes, given the patient's history up to that stage. This assumption is crucial for ensuring unbiased estimation of treatment effects.
     \item \( A.3 \): Positivity requires a positive probability of any treatment allocation, i.e., for any $a_j \in \mathcal{A}_j$ and  $1 \leq j \leq K,$ we have  $\mathbb{P}(A_j=a_j \mid \mathbf{h}_j) > 0$.
          \item \( A.4 \): Consistency means that the potential outcome under a given sequence of treatments is equal to the observed outcome if, in fact, those treatments were the ones actually received.
\end{itemize}

We define the (expected) outcome model given the full history matrix $\mathbf{h}$ as
\begin{eqnarray}\label{equ:EOM}
 \mathbb{E}(Y^{\overline{\mathbf{a}}_K}|\mathbf{h}; \pmb{\beta}, \pmb{\Psi}) & = & \mathbb{E}(Y^{\overline{\mathbf{0}}_{K}}|\mathbf{h}^{\beta}; \pmb{\beta})\nonumber\\
   & + &   \sum_{j=1}^{K} \mathbb{E}(Y^{\overline{\mathbf{a}}_j, \underline{\mathbf{a}}_{j+1}^{opt}}-Y^{\overline{\mathbf{a}}_{j-1},0, \underline{\mathbf{a}}_{j+1}^{opt}}|\mathbf{h}_j^{\Psi};\pmb{\Psi}_j),\\\label{equ:tfm+bilp}
   & = & \sum_{j=1}^{K}  \left[ f_j(\mathbf{h}_j^{\beta}; \pmb{\beta}_j)+ \gamma_j(\mathbf{h}_j^{\Psi}, a_j; \pmb{\Psi}_j) \right],
\end{eqnarray}
where $\mathbf{h}_j^{\beta}$ and $\mathbf{h}_j^{\Psi}$ are subsets of the patient's history matrix $\mathbf{h}_j$  available prior to the $j$th treatment decision. The matrix $\mathbf{h}_j^{\beta}$ includes non-treatment (predictor) covariates, while $\mathbf{h}_j^{\Psi}$ contains tailoring (prescriptive) covariates. The relation \eqref{equ:EOM} represents the expected outcome for a patient with history $\mathbf{h}$ who follows the treatment regime $\overline{\mathbf{a}}_K$.

 In the DTR literature, equation \eqref{equ:EOM} is known as the \textit{outcome model}, while the summands on the right-hand side of equation \eqref{equ:tfm+bilp} are referred to as the \textit{treatment-free model} parameterized by $\pmb{\beta}_j$ ($f_j(\mathbf{h}_j^{\beta}; \pmb{\beta}_j)$) and the \textit{blip function} at stage $j$ parameterized by $\pmb{\Psi}_j$ ($\gamma_j(\mathbf{h}_j^{\Psi}, a_j; \pmb{\Psi}_j))$, respectively.  This expresses the expected outcome for a patient with history $\mathbf{h}$ who receives the treatment regime $\overline{\mathbf{a}}_K$ in terms of two components: first, the treatment-free model, which reflects the expected outcome if no treatments $(\overline{\mathbf{0}}_K = (0, \ldots, 0)$) are administered throughout the regime, and is affected solely by non-treatment covariates at each stage stored in $\mathbf{h}_j^{\beta}$. The second component accounts for the effect of the treatments received through the regime, which is quantified at each stage by the blip function $\gamma_j$ as
 \begin{eqnarray*}
    \gamma_j(\mathbf{h}_j^{\Psi}, a_j; \pmb{\Psi}_j)= \mathbb{E}(Y^{\overline{\mathbf{a}}_j, \underline{\mathbf{a}}_{j+1}^{opt}}-Y^{\overline{\mathbf{a}}_{j-1},0, \underline{\mathbf{a}}_{j+1}^{opt}}|\mathbf{h}_j^{\Psi};\pmb{\Psi}_j).
\end{eqnarray*}

More precisely, these blip functions measure the additional impact of receiving the treatment $a_j$, compared to not receiving the treatment ($a_j = 0$), for a patient with a similar history who is expected to receive the optimal (not necessarily identical) treatments in the subsequent stages. It can be understood from the definition that the blip function under the no-treatment scenario is zero, i.e.,  $\gamma_j(\mathbf{h}_j^{\Psi}, 0; \pmb{\Psi}_j)=0$.

 Another concept related to the dWOLS setting is the regret (loss) function at stage $j$ defined as 
\begin{align*}
\mu_j(\mathbf{h}_j^{\Psi}, a_j; \pmb{\Psi}_j)=\mathbb{E}(Y^{\overline{\mathbf{a}}_{j-1}, \underline{\mathbf{a}}_{j}^{opt}}-Y^{\overline{\mathbf{a}}_{j}, \underline{\mathbf{a}}_{j+1}^{opt}}|\mathbf{h}_j^{\Psi};\pmb{\Psi}_j),
\end{align*}
that quantifies the expected loss incurred by prescribing treatment $a_j$ at stage $j$ rather than the optimal treatment denoted by $a_j^{opt}$, assuming optimal treatments are received thereafter.  In the current problem setting, the blip and regret functions are related by
\begin{align*}
 \mu_j(\mathbf{h}_j^{\Psi}, a_j; \pmb{\Psi}_j)=\gamma_j(\mathbf{h}_j^{\Psi}, a^{opt}_j; \pmb{\Psi}_j)-\gamma_j(\mathbf{h}_j^{\Psi}, a_j; \pmb{\Psi}_j). 
\end{align*}

The regret affords an alternate formulation of the outcome model as
\begin{align}\label{equ:EOMR}
\mathbb{E}(Y^{\overline{\mathbf{a}}_K}|\mathbf{h}; \pmb{\Psi})&= \mathbb{E}(Y^{opt}|\mathbf{h})- \sum_{j=1}^{K}  \mu_j(\mathbf{h}_j^{\Psi}, a_j; \pmb{\Psi}_j),
\end{align}
where $Y^{opt}$ is the (theoretical) outcome under the optimal DTR. The relation \eqref{equ:EOMR} states that the potential outcome equals the theoretically optimal (the best possible) outcome minus the cumulative negative effects resulting from any suboptimal treatment decisions. Therefore, if all treatment decisions in the prescribed regime are optimal, the total regret will be zero since $\mu_j(\mathbf{h}_j^{\Psi}, a_j^{opt}; \pmb{\Psi}_j)=0$ for $j=1,\ldots,K$, and the observed outcome will match the optimal one.

\begin{figure}
\center
\includegraphics[scale=0.3]{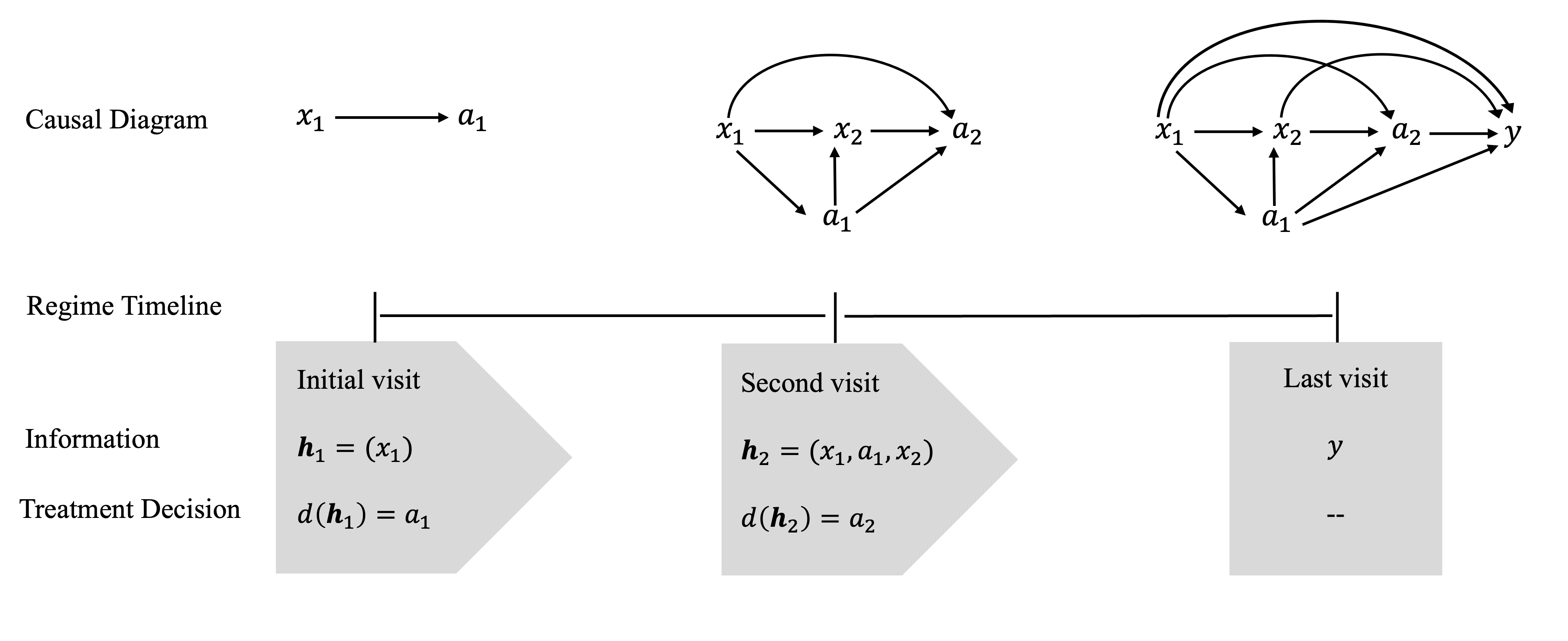}
\caption{Schematic of a two-stage treatment regime with one covariate along with causal diagrams at each stage.}
\end{figure}\label{fig:CD}

The optimal DTR is the sequence of treatment decisions that maximizes the expected clinical benefits for each individual based on their information. The outcome model in \eqref{equ:EOM} and \eqref{equ:tfm+bilp} states that the treatment decisions influence the outcome only through the stage-specific blips $\gamma_j$. Mathematically, the stage $j$ treatment decision can be defined as $a_j=d(\mathbf{h}_j^{\Psi}): \mathbf{h}_j^{\Psi} \rightarrow \mathcal{A}_j$. Morevere, the optimal treatment at stage $j$ is that which maximizes the blip, i.e., $a^{opt}_j = \underset{a_j \in \mathcal{A}_j}{\arg\max} \, \gamma_j(\mathbf{h}_j^{\Psi}, a_j; \pmb{\Psi}_j)$, or equivalently, that which minimizes the regret, i.e., $a^{opt}_j = \underset{a_j \in \mathcal{A}_j}{\arg\min} \, \mu_j(\mathbf{h}_j^{\Psi}, a_j; \pmb{\Psi}_j)$.

To determine $a^{opt}_j$, it is necessary to specify the mathematical forms of the treatment-free model $f$ and the blip $\gamma$ at each stage. In the related literature, linear models are commonly recommended for both functions. Specifically, the treatment-free outcome model is typically specified as $f_j(\mathbf{h}_j^{\beta}; \pmb{\beta}_j) = \pmb{\beta}_j^\top \mathbf{h}_j^{\beta}$, and the blip function is modeled as $\gamma_j(\mathbf{h}_j^{\Psi}, a_j; \pmb{\Psi}_j) = \pmb{\Psi}_j^\top a_j \mathbf{h}_j^{\Psi}$ where the superscript $\top$ stands for transposition. Accordingly, if one estimates $\pmb{\Psi}_j$, denoting this estimate by $\widehat{\pmb{\Psi}}_j$, then $a_j=1$ is the optimal stage $j$ treatment if $\widehat{\pmb{\Psi}}_j^{T}\mathbf{h}_j^{\Psi} > 0$; otherwise the control ($a_j=0$) is the optimal decision. Using these linear models, for instance, for a one-stage regime ($K=1$) with a single covariate $x$, the outcome model in \eqref{equ:EOM} and \eqref{equ:tfm+bilp} reduces to $E(Y^a|x,\beta_0,\beta_1,\psi_0,\psi_1)=\beta_0+\beta_1x+a(\psi_0+\psi_1x)$. Then, $a_j=1$ is the optimal treatment for a patient with covariate $x$ if $\hat{\psi}_0+\hat{\psi}_1x >0$. Figure \ref{fig:CD} shows the schematic of a two-stage treatment regime with one covariate along with causal diagrams at each stage.

\subsection*{Balancing weights}

To estimate the parameters $\pmb{\beta}_j$ and $\pmb{\Psi}_j$, one must correctly specify both the treatment-free model as well as the blip if we wish to have consistent estimators. In practice, the treatment-free model, in particular, may be difficult to specify, and so an approach that affords robustness to its misspecification is desirable. The dWOLS approach tackles this obstacle by introducing the \textit{treatment model} for the relationship between treatment assignment and covariates. The treatment model can be expressed as $\pi(\mathbf{h};\pmb{\alpha})=\mathbb{P}(A=1|\mathbf{h};\pmb{\alpha})$, the probability of receiving treatment given the history $\mathbf{h}$ parameterized by $\pmb{\alpha}$. The treatment model accounts for potential confounding in treatment assignment and is typically approached using logistic regression, i.e., $\pi(\mathbf{h};\pmb{\alpha})=\frac{1}{1 + \exp\left(-\pmb{\alpha}^\top \mathbf{h}\right)}$. To estimate the blip parameters, dWOLS uses weighted ordinary least squares estimation based on the weights $w(\mathbf{h}, a)$ that satisfy the condition
\begin{align}\label{equ:balweight}
\pi(\mathbf{h};\pmb{\alpha})w(\mathbf{h}, 1)=(1-\pi(\mathbf{h};\pmb{\alpha}))w(\mathbf{h}, 0).
\end{align}

Using weighted data in estimation helps balance covariate distributions across treatment groups. This approach enables a more accurate estimation of treatment effects, particularly in scenarios like observational data, where treatment assignments are influenced by confounding factors.

\subsection*{Implementation procedure}

In the multi-stage setting, dWOLS employs a recursive estimation approach, beginning at the final stage of treatment and moving backward to earlier stages. This approach allows for the estimation of treatment effects while accounting for future treatments.  The process is detailed in the following two key steps
\begin{itemize}
    \item[Step 1] (Estimation at the final stage $j = K$)
    Set $\tilde{y}_K$ to be to the observed outcome $y$, i.e., $\tilde{y}_K = y$.   Estimate the treatment model parameter $\pmb{\alpha}_K$. Then, perform a weighted ordinary least squares regression of $\tilde{y}_K$ on the stage $K$ history $\mathbf{h}_K^{\beta}$ and $a_K \mathbf{h}_K^{\Psi}$ using the weights $w_K(\mathbf{h}_K;\pmb{\hat{\alpha}}_K)$ to obtain the blip parameter estimate $\pmb{\hat{\Psi}}_K$.
    \item[Step 2] (Recursive estimation) 
    At each earlier stage $j < K$, calculate the pseudo-outcome $\tilde{y}_j$ by
    \begin{align}\label{pseudo-outcome}
     \tilde{y}_j = y + \sum_{k=j+1}^K \mu_k(\mathbf{h}_k^{\Psi}, a_k; \hat{\pmb{\Psi}}_k).
     \end{align}
      Then, in the same manner as in stage $K$, perform a weighted ordinary least squares regression of $\tilde{y}_j$ on the up-to-stage-$j$ history $\mathbf{h}_j^{\beta}$ and $a_j \mathbf{h}_j^{\Psi}$ using the weights $w_j(\mathbf{h}_j;\pmb{\hat{\alpha}}_j)$ to obtaine the blip parameter estimate $\pmb{\hat{\Psi}}_j$.   
\end{itemize}
 
 The pseudo-outcome $\tilde{y}_j$ in \eqref{pseudo-outcome} can be thought of as the expected outcome of a patient who receives their particular regime up to stage $j$ and then is optimally treated afterward. Wallace and Moodie \cite{wallace2015doubly} demonstrated that the dWOLS method yields consistent estimators of the blip parameters if at least one of the treatment or treatment-free models is correctly specified, a property known as \textit{double robustness}. They recommend using the weight $w(\mathbf{h}; \pmb{\alpha}) = \left| a - E(A \mid\mathbf{h}; \pmb{\alpha}) \right|$, which assigns more weight to patients who receive less-expected treatments based on their health information.  The principle behind double robustness is that if the treatment model is correct, the weighted data approximates a randomized experiment, allowing for consistent treatment effect estimation. If the treatment model is incorrect, controlling for confounders in the outcome model through a correctly specified treatment-free model still helps to reduce bias.

  To implement dWOLS, one therefore needs to specify the form of the treatment model $\pi(\mathbf{h};\pmb{\alpha})$, the weight $w(\mathbf{h}, a)$, and the pseudo-outcome model at each stage. While suggestions have been presented for the first two models, the dWOLS approach addresses the last concern by assuming
\begin{align*} 
E(\tilde{Y}_j^{a_j}|\mathbf{h}_j; \pmb{\beta}_j, \pmb{\Psi}_j)&=\pmb{\beta}_j^\top \mathbf{h}_j^{\beta} + \pmb{\Psi}_j^\top a_j \mathbf{h}_j^{\Psi} \, \text{for $j=1,\ldots,K$},
\end{align*}
where $\tilde{Y}_K=Y$.  To further support practitioners, Wallace et al. \cite{wallace2016package,wallace2017dynamic} introduced the DTRreg R package, which facilitates the application of dWOLS (and some of its extensions).

We conclude this subsection by noting that, moving forward, we will aim to maintain consistency with the standard dWOLS notation throughout the paper (e.g., $Y, a, \mathbf{h}$, etc.). New notations will only be introduced when necessary. Table \ref{tab:notations} shows the notations used in the paper. Additionally, many concepts—such as potential outcomes, blip, regret, and pseudo-outcomes—will be used with the same interpretation as in the dWOLS framework, and thus will not be elaborated upon further.

\subsection*{Applications}

The dWOLS approach was first illustrated using data from the Promotion of Breastfeeding Intervention Trial, which involved 17,046 mother-infant pairs in Belarus who were randomized to either a breastfeeding promotion intervention or standard care. Participants were followed across seven time points up to 12 months, with various measurements recorded, including infant weight and breastfeeding status, forming a three-stage dynamic treatment regime. The primary covariate used for tailoring treatment decisions was the infant’s weight, and the main outcome of interest was the infant’s weight at 12 months. The objective of DTR estimation in this example was to optimize infant growth by recommending stage-specific breastfeeding decisions based on the evolving weight of the infant.

\section{Extensions of dWOLS}
\label{sec:Ext}

The standard dWOLS method assumes a continuous outcome of interest and a binary treatment decision at each stage. However, many clinical studies involve different scenarios, such as binary or survival outcomes. Additionally, treatment decisions may require selecting from more than two options or determining an individualized drug dosage. This section reviews studies that have extended the dWOLS framework to accommodate a wider range of outcome types and treatment options.

 \subsection{G-dWOLS: dWOLS with continuous treatment}
 \label{sec:G-dWOLS}
 
 There are practical situations in which treatment decisions involve tailoring continuous dosages to individual patient needs rather than selecting between two options. This challenge is commonly referred to as dose finding in the related literature \cite{rich2016optimal}. Schulz and Moodie \cite{schulz2021doubly} extended the dWOLS approach by introducing the generalized dWOLS (G-dWOLS) method to apply to the class of continuous treatments.
 
 \subsection*{Notations and blip model}
 
The treatment space $\mathcal{A}$ is a continuous set in the G-dWOLS setting. The outcome of interest here is still the continuous variable $Y$. The G-dWOLS framework closely resembles the dWOLS approach, with some key generalizations and modifications. To determine the treatment $a \in \mathcal{A}$ that maximizes the blip, the blip function should typically be parameterized to ensure that the optimal treatment could be an interior point of $\mathcal{A}$. For example, the blip model might include a quadratic term for the treatment, i.e.,
\begin{align}\label{equ:blipgdWOLS}
\gamma(\mathbf{h}^{\Psi}, a; \pmb{\Psi}) = a \pmb{\Psi}^{(1)\top} \mathbf{h}^{\Psi(1)} + a^2 \pmb{\Psi}^{(2)\top} \mathbf{h}^{\Psi(2)},
 \end{align}
where $\pmb{\Psi}^{(1)}$ and $\pmb{\Psi}^{(2)}$ are the first-order and second-order interaction coefficients with respect to the treatment variable $a$.

According to \eqref{equ:blipgdWOLS}, whenever  $\pmb{\Psi}^{(2)\top} \mathbf{h}^{\Psi(2)} < 0$,  the optimal dosage is then $a^{opt} = -\frac{\pmb{\Psi}^{(1)\top} \mathbf{h}^{\pmb{\Psi}(1)}}{2 \pmb{\Psi}^{(2)\top} \mathbf{h}^{\pmb{\Psi}(2)}}$. Thus, as in the dWOLS framework, determining the optimal DTR here simplifies to the problem of estimating the blip parameters $\pmb{\Psi}^{(1)}$ and $\pmb{\Psi}^{(2)}$.

\subsection*{Balancing weights and implementation procedure} 
 Analogous to the treatment model in the original dWOLS, Schulz and Moodie \cite{schulz2021doubly} defined the generalized propensity score as the conditional probability density function $\pi(\mathbf{h}, a; \pmb{\alpha})=f_{A|\mathbf{h}}(a|\mathbf{h}; \pmb{\alpha})$ for any $a \in A$ and $\pi(a; \pmb{\alpha})$ as the corresponding null propensity score that does not account for the covariate effects. They proved that weighted ordinary least square regressions yield consistent estimators of the blip parameters as long as at least either the treatment-free model or treatment model is correctly specified. To have this, the weight $w(\mathbf{h}, a)$ should satisfy the condition $\pi(\mathbf{h}, a; \pmb{\alpha}) w(\mathbf{h},a)=\pi(\mathbf{h}, a^{'}; \pmb{\alpha})w(\mathbf{h}, a^{'})$ for any $a \neq a^{'} \in \mathcal{A}$. This condition is simply the extension of the balancing condition in the original dWOLS approach given in \eqref{equ:balweight}. Schulz and Moodie \cite{schulz2021doubly} suggested a wide family of weights, including the inverse probability of treatment weights (IPW) $w(\mathbf{h}, a)=\frac{1}{\pi(\mathbf{h}, a; \pmb{\alpha})}$ and stabilized IPW (SIPW) $w(\mathbf{h},  a)=\frac{\pi(a;  \pmb{\alpha})}{\pi(\mathbf{h}, a; \pmb{\alpha})}$ where $\pi(a; \pmb{\alpha})$ follows a uniform (either continuous or discrete) distribution. The implementation procedure of the G-dWOLS is also the same as dWOLS, starting with estimation of stage $K$ parameters and then moving backwards to estimate the stage $j=1,\ldots, K-1$ parameters using the pseudo-outcome as in \eqref{pseudo-outcome}.

\subsection*{Applications}

 An example of continuous dosing is warfarin, an anticoagulant where the dose (in mg/week) is adjusted based on the patient's international normalized ratio (INR) to achieve a therapeutic range, typically varying from 1 mg to over 10 mg per week depending on individual factors such as age, diet, and genetics. Schulz and Moodie \cite{schulz2021doubly} illustrated the G-dWOLS approach to estimate an optimal warfarin dosing to treat and prevent thrombosis in a single-stage setting. They also used patient-related information, including demographic and genetic factors, to maximize a clinical outcome related to INR levels. They used the simple transformation to redefine the outcome as a larger-the-better variable as $Y=-\sqrt{|2.5-\text{INR}|}$.

Ma et al. \cite{ma2021individualized} utilized the G-dWOLS estimation approach to optimize treatment strategies for septic shock, focusing on fluid volume intake and norepinephrine dosing as continuous treatments. By applying G-dWOLS, the study aims to identify the optimal dosing strategies that maximize survival outcomes, assessed via mortality probability as a continuous variable. Their results suggest that using DTR estimation to adapt fluid and norepinephrine dosages according to patient information can potentially improve survival outcomes in septic shock management.

Hong et al. \cite{hong2021individualized} applied G-dWOLS to optimize mechanical ventilation (MV) strategies for patients with acute respiratory failure (ARF), focusing on mechanical power (MP) as a continuous treatment. The study aimed to identify individualized MP dosing that improves mortality risk, assessed over the initial 48 hours of MV. The optimal MP values were estimated and compared with actual MP levels. Results indicated that adjusting MP doses according to optimal values derived through DTR could improve mortality outcomes in ARF patients.

\subsection{dWGLM: dWOLS with binary outcome}
\label{sec:dWGLM}
Jiang et al. \cite{jiang2022doubly} extended the original dWOLS framework to accommodate binary outcomes by proposing the dynamic weighted generalized linear model (dWGLM), while the treatment option at each stage is still binary.  

\subsection*{Notations and blip model}

Let $Y$ denote the binary outcome for an individual where $Y = 1$ represents a desirable outcome. As in the original dWOLS methodology, the concepts of blip and regret are central to the analysis and are mathematically defined at stage $j$ as
\begin{eqnarray*}
\gamma_j(\mathbf{h}_j^{\Psi}, a_j; \pmb{\Psi}_j) &=& g \left( \mathbb{P} [Y^{\overline{\mathbf{a}}_j, \overline{\mathbf{a}}_{j+1}^{opt}} = 1 \mid \mathbf{h}_j^{\Psi} ]; \pmb{\Psi}_j\right)\\
&-& g \left( \mathbb{P} [Y^{\overline{\mathbf{a}}_{j-1}, 0, \overline{\mathbf{a}}_{j+1}^{opt}} = 1 \mid \mathbf{h}_j^{\Psi}]; \pmb{\Psi}_j\right),\\
\mu_j(\mathbf{h}_j^{\Psi}, a_j; \pmb{\Psi}_j) &=& g \left( \mathbb{P} [Y^{\overline{\mathbf{a}}_{j-1}, \overline{\mathbf{a}}_{j}^{opt}} = 1 \mid \mathbf{h}_j^{\Psi}]; \pmb{\Psi}_j\right)\\
&-& g \left( \mathbb{P} [Y^{\overline{\mathbf{a}}_j, \overline{\mathbf{a}}_{j+1}^{opt}} = 1 \mid \mathbf{h}_j^{\Psi}]; \pmb{\Psi}_j\right)\\
&=& \gamma_j(\mathbf{h}_j^{\Psi}, a_j^{opt}; \pmb{\Psi}_j) - \gamma_j(\mathbf{h}_j^{\Psi}, a_j; \pmb{\Psi}_j),
\end{eqnarray*}
where $g: (0,1) \rightarrow (-\infty, \infty)$ is the link function.

Common choices for the link function include the identity link $g(p) = p$, the logit function $g(p) = \log\left(\frac{p}{1-p}\right)$, and the probit link $g(p) = \Phi^{-1}(p)$, where $\Phi$ denotes the cumulative distribution function of the standard normal distribution. Building on the concepts of blip and regret, the dWGLM approach introduces two outcome probabilities that parallel the outcome model in dWOLS (defined in \eqref{equ:EOM} and \eqref{equ:EOMR}) as
\begin{align}\label{equ:EOMB}
 g \left(\mathbb{P} [Y^{\overline{\mathbf{a}}_K}=1|\mathbf{h}]; \pmb{\beta}, \pmb{\Psi}\right) &=\sum_{j=1}^{K} \left[ f_j(\mathbf{h}_j^{\beta}; \pmb{\beta}_j)+ \gamma_j(\mathbf{h}_j^{\Psi}, a_j; \pmb{\Psi}_j)\right],\nonumber\\
 &=g \left(\mathbb{P}[Y^{\overline{\mathbf{a}}_K^{opt}}=1|\mathbf{h}]; \pmb{\beta}, \pmb{\Psi}\right)\\
 &- \sum_{j=1}^{K}  \mu_j(\mathbf{h}_j^{\Psi}, a_j; \pmb{\Psi}_j)\nonumber.
\end{align}

According to \eqref{equ:EOMB}, the treatment regime $\overline{\mathbf{a}}_K$ is optimal if it maximizes $g \left(\mathbb{P} [Y^{\overline{\mathbf{a}}_K} = 1 \mid \mathbf{h}]; \pmb{\beta}, \pmb{\Psi}\right)$, which is equivalent to maximizing the blip function at each stage, given the patient’s history. Thus, with the estimates $\pmb{\hat{\Psi}}_j$ for $j = 1, \ldots, K$, prescribe $a_j = 1$ if $\gamma_j(\mathbf{h}_j^{\Psi}, a_j; \pmb{\hat{\Psi}}_j) > 0$; otherwise, prescribe no treatment.

\subsection*{Balancing weights}

 The main aim hereafter is to estimate the blip parameters for binary outcomes in a multi-stage treatment decision problem based on the above discussion. This is achieved by employing a series of weighted GLMs to estimate the parameters at each stage. The dWGLM method suggests using linear models to relate the matrices $\mathbf{h}_j^{\beta}$ and $\mathbf{h}_j^{\Psi}$ to the outcome probability, i.e., $f_j(\mathbf{h}_j^{\beta}; \pmb{\beta})=\pmb{\beta}_j^\top \mathbf{h}_j^\beta$ and $\gamma_j(\mathbf{h}_j^{\Psi}, a_j; \pmb{\Psi}_j)=\pmb{\psi}_j^\top a_j \mathbf{h}_j^{\Psi}$. Jiang et al. \cite{jiang2022doubly} proved that weighted GLMs will give consistent estimates of the blip if the weight satisfy
\begin{eqnarray}\label{equ:weigth}
&&\pi(\mathbf{h}; \pmb{\alpha})w(\mathbf{h}, 1)k(\mathbf{h}, 1)=\\
&& \qquad \qquad \qquad \qquad (1-\pi(\mathbf{h}; \pmb{\alpha}))w(\mathbf{h}, 0)k(\mathbf{h}, 0)\nonumber,
\end{eqnarray}
 where $\pi(\mathbf{h}; \pmb{\alpha})$ is the treatment model and $k(\mathbf{h}, a)$ is given in \textbf{Theorem 1} of Jiang et al. (2022).

 \subsection*{Implementation procedure}

  A weighted GLM that utilizes weights satisfying the condition in \eqref{equ:weigth} will ensure doubly robust estimates of the blip parameter $\pmb{\Psi}_j$ for $j=1,\ldots,K$. However, the condition in \eqref{equ:weigth} necessitates a two-step process for estimating the balancing weights. First, the term $k(\mathbf{h}, a)$ must be estimated using a weighted GLM with dWOLS weights, which satisfy the equation $\pi(\mathbf{h};\pmb{\alpha})w(\mathbf{h}, 1) = (1-\pi(\mathbf{h};\pmb{\alpha}))w(\mathbf{h}, 0)$. Once $k(\mathbf{h}, a)$ is estimated, the weights $w^{new}(\mathbf{h}, a)$ for the dWGLM can be calculated accordingly so that the condition in \eqref{equ:weigth} holds. Consequently, the dWGLM method can be applied by using the following steps at each stage.

\begin{enumerate}
    \item[Step 1]  If $j = K$, set $\tilde{y}_j = y$. Otherwise, generate $R$ pseudo outcomes $\tilde{y}_j^1, \ldots, \tilde{y}_j^R$ representing $\tilde{Y}_j$ from a Bernoulli distribution with the success probability
\begin{eqnarray*}
     \mathbb{P}(\tilde{Y}_j = 1) &=& g^{-1}\Big[g\left(\mathbb{P}(Y = 1 \mid \mathbf{h}_K, a_K; \hat{\pmb{\beta}}_K, \hat{\pmb{\Psi}}_K)\right)\\
    && + \sum_{k=j+1}^K \mu_k(\mathbf{h}_k^{\Psi}, a_k; \hat{\pmb{\Psi}}_k)\Big].
\end{eqnarray*}
    \item[Step 2] Estimate the treatment model $\pi_j(\mathbf{h};\pmb{\alpha})$, compute the weight $w_j(\mathbf{h}, a)$, and then perform a weighted GLM of \( \tilde{y}_j^r \) on $\pmb{\beta}_j^\top \mathbf{h}_j^\beta$ and $\pmb{\psi}_j^\top a_j \mathbf{h}_j^{\Psi}$ using this weight to get estimates $\hat{\pmb{\beta}}^{\text{old},r}, \hat{\pmb{\psi}}_j^{\text{old},r}$ for $r = 1, \ldots, R$.
    \item[Step 3] Given the estimates in Step 2, calculate $ k_j^r(\mathbf{h}, a_j)$ for $r=1,\ldots,R$. Then, construct the new weights from $w_j^{\text{new},r}(\mathbf{h}, a_j) = w_j(\mathbf{h},a) \times k_j^r(\mathbf{h}, 1 - a_j)$.  Perform a weighted GLM with the new weights $w_j^{\text{new},r}(\mathbf{h}_j, a_j)$ to get revised estimates $(\hat{\pmb{\beta}}_j, \hat{\pmb{\Psi}}_j^r)$ for $r=1,\ldots,R$. Eventually, estimate $\pmb{\Psi}_j \) by \(\hat{\pmb{\Psi}}_j = \frac{1}{R} \sum_{r=1}^R \hat{\pmb{\Psi}}_j^r$.
\end{enumerate}

\subsection*{Application}

The dWGLM approach has been illustrated using data from the Population Assessment of Tobacco and Health (PATH) study, focusing on a three-stage dynamic treatment regime for smoking cessation. The binary outcome in this study is whether an individual quits smoking and the binary treatment is the use or non-use of e-cigarettes. At each stage, covariates for each patient include age, education, sex, non-Hispanic race, and their plan to quit smoking. The analysis aims to identify the optimal sequence of treatment decisions over these three stages to maximize the probability of smoking cessation.

\subsection{DWSurv: dWOLS with survival outcomes}
\label{sec:DWSurv}

The dynamic weighted survival modeling (DWSurv) method developed by Simoneau et al. \cite{simoneau2020estimating} extends the dWOLS approach to accommodate survival outcomes that are subject to right-censoring, while the treatment decisions are still binary. The goal of this method is to identify a sequence of treatment decisions that maximize the expected survival time for each individual.

\subsection*{Notations and blip model}
Let $\eta_{j}$ be an indicator variable where $\eta_{j} = 1$ denotes that an individual has entered stage $j$, and $\eta_{j} = 0$ otherwise. All individuals are assumed to enter at least the first stage, i.e., $\eta_{1} = 1$. At each stage, the outcome of interest is the survival time within that stage, denoted by $Y_j$, where $Y_j = 0$ if $\eta_{j} = 0$. Consequently, the individual's overall survival time, denoted by $Y$, is defined as $Y = \sum_{j=1}^K \eta_j Y_j$. In the presence of the treatment regime $\overline{\mathbf{a}}_K$, the counterfactual survival time is represented by $Y^{\overline{\mathbf{a}}_K}$, and is defined as $Y^{\overline{\mathbf{a}}_K} = \sum_{j=1}^K \eta_j Y_j^{\overline{\mathbf{a}}_j}$. Additionally, the censoring time is denoted by $C$, and the observed survival time is $Y^c = \min\{Y, C\}$. The censoring indicator $\Delta$ is defined such that $\Delta = 0$ indicates that the survival time is censored. For illustration, we present the case of two-stage treatments ($K=2$), and the case $K > 2 $ follows similarly.

 Similar to the dWOLS, the DWSurv approach begins with the last stage $K=2$ and moves backward to estimate the decision parameters in stage one. The stage 2 blip function is defined as
\begin{eqnarray}\label{equ:blipDWSurv2}
&&\gamma_2(\mathbf{h}_2^{\Psi}, a_2; \pmb{\Psi}_2)=\\
&&\qquad \, \mathbb{E}\left(\log(Y^{a_1,a_2})-\log(Y^{a_1,0})|\eta_2=1, \mathbf{h}_2^{\Psi};\pmb{\Psi}_2\right)\nonumber,
\end{eqnarray}
so that $\gamma_2(\mathbf{h}_2^{\Psi}, 0; \pmb{\Psi}_2)=0$. According to \eqref{equ:blipDWSurv2}, we only use data from individuals who entered the second stage to estimate the decision parameter $\pmb{\Psi}_2$. Thus, given the information stored in $\mathbf{h}_2^{\Psi}$, the optimal second stage treatment $a_2^{opt}$ can be determined by $a^{opt}_2 = \underset{a_2 \in \{0,1\}}{\arg\max} \, \gamma_2(\mathbf{h}_2^{\Psi}, a_2; \pmb{\Psi}_2)$ which would lead to the maximum expected survival time within stage $K=2$.

Given the optimal stage 2 treatment $a^{opt}_2$, the first stage blip model is
\begin{eqnarray}\label{equ:blipDWSurv1}
&& \gamma_1(\mathbf{h}_1^{\Psi}, a_1; \pmb{\Psi}_1)=\\
&& \qquad \quad \mathbb{E}\left(\log(Y^{a_1,a_2^{opt}})-\log(Y^{0,a_2^{opt}})| \mathbf{h}_1^{\Psi};\pmb{\Psi}_1\right)\nonumber,
\end{eqnarray}
where $Y^{a_1,a_2^{opt}}=Y_1^{a_1}+\eta_2 Y_2^{a_1,a_2^{opt}}$ is the pseudo-overall survival time had an individual received their optimal stage 2 treatment given the information that they entered stage 2. By this definition, individuals who did not enter the second stage have their pseudo-outcome equal to their overall survival time, which is equal to the time $Y_1^{a_1}$ spent in the first stage. 

To calculate the expectation in the blip functions \eqref{equ:blipDWSurv2} and \eqref{equ:blipDWSurv1}, we define outcome models for each stage that relate the log-survival time of each stage to all available information from that stage. Simoneau et al. \cite{simoneau2020estimating} proposed using the well-known accelerated failure time (AFT) model along with linear functions, though non-linear functions may also be accommodated. Following this approach, for stage 2 we have
\begin{align}\label{equ:outMo}
\log(Y_2^{a_1,a_2})=\pmb{\beta}_2^\top \mathbf{h}_2^{\beta} + \pmb{\Psi}_2^\top a_2 \mathbf{h}_2^{\Psi}+\epsilon_2,
\end{align}
where $\epsilon_2$ are \textit{iid} random variables (across individuals) with $\mathbb{E}(\epsilon_2=0)$. Given \eqref{equ:blipDWSurv2} and \eqref{equ:outMo} and the fact that $\mathbb{E}(\epsilon_2=0)$, it is easy to see that $\gamma_1(\mathbf{h}_2^{\Psi}, a_2; \pmb{\Psi}_2)=\pmb{\Psi}_2^\top a_j \mathbf{h}_2^{\Psi}$. In the same manner as \eqref{equ:outMo}, we define 
\begin{align*}
\log(Y^{a_1,a_2^{opt}})=\pmb{\beta}_1^\top \mathbf{h}_1^{\beta} + \pmb{\Psi}_1^\top a_1 \mathbf{h}_1^{\Psi}+\epsilon_1.
\end{align*}

Though $Y_2^{a_1,a_2}$ is observable, $Y_2^{a_1,a_2^{opt}}$ is not. This means that to estimate the blip parameter $\pmb{\Psi}_1$, we need to generate the pseudo outcomes. Simoneau et al. \cite{simoneau2020estimating} suggested the following pseudo outcome  
\begin{eqnarray}\label{equ:psudo}
&&\tilde{Y}^{a_1,a_2^{opt}}=Y_1^{a_1}\\
&& \qquad \quad +\eta_2\left(Y_2^{a_1,a_2} \times \exp(\pmb{\Psi}_2^\top \mathbf{h}_2^{\Psi}[a_2^{opt}-a_2])\right).\nonumber
\end{eqnarray}

The relation \eqref{equ:psudo} constructs counterfactual survival times under $a^{opt}_2$ by adding a positive quantity to the observed survival times of the individuals who did not receive their optimal stage 2 treatment. The first stage treatment comparison is then ``fair'' as it is with respect to an overall survival time that incorporates the effect of the stage 2 treatment, taken to be optimal for everybody. Thus, an individual who received their optimal stage 2 treatment has the term inside the $\exp(\cdot)$ equal to zero and the pseudo-outcome equal to the observed survival time, that is, $\tilde{Y}^{a_1,a_2^{opt}}$. An individual who did not receive their optimal treatment has the term inside the $\exp(·)$ greater than zero and a pseudo-outcome larger than the observed outcome, that is, $\tilde{Y}^{a_1,a_2^{opt}} > Y$. An individual who did not enter the second stage has a pseudo-outcome equal to the observed outcome, i.e., $\tilde{Y}^{a_1,a_2^{opt}}=Y_1^{a_1}$. Eventually, the log-outcome models given in \eqref{equ:outMo} and \eqref{equ:psudo} can be used to estimate the blip parameters $\pmb{\Psi}_1$ and $\pmb{\Psi}_2$.

\subsection*{Balancing weights}

 \textbf{Theorem 1} of Simoneau et al. \cite{simoneau2020estimating} proved that a weighted generalized estimation equation (GEE) approach ensures doubly robust estimates of the blip parameters if the weights $w(\mathbf{h}, a, \delta)$ satisfy
\begin{eqnarray}\label{equ:BW}
 &&g(\mathbf{h}, 1)\pi(\mathbf{h}; \pmb{\alpha})w(\mathbf{h}, 1,1)=\\
 && \qquad \qquad (1-g(\mathbf{h}, 1))(1-\pi(\mathbf{h}; \pmb{\alpha}))w(\mathbf{h}, 0,0)\nonumber,
\end{eqnarray}
where $\pi(\mathbf{h}; \pmb{\alpha})=\mathbb{P}(A=1|\mathbf{h}, \eta=1; \pmb{\alpha})$ is the treatment model and  $g(\mathbf{h}, a)=\mathbb{P}(\Delta=1|\mathbf{h}, a, \eta=1)$ is the censoring model. By extending the weights of the standard dWOLS but in the presence of censored data, one option for balancing weights is $w(\mathbf{h}, a, \delta)=\frac{|a-\mathbb{P}(A=1|\mathbf{h}, \eta=1; \pmb{\alpha})|}{\mathbb{P}(\Delta=\delta|\mathbf{h}, a, \eta=1)}$.

\subsection*{Implementation procedure}
The following steps can be applied to estimate the optimal DTR for a two-stage treatment regime. Before this, the treatment model $\pi(\mathbf{h}; \pmb{\alpha})$ and the censoring model $g(\mathbf{h}, a)$ at each stage should be specified.

\begin{enumerate}
    \item Specify weights $w_2(\mathbf{h}_2, a_2, \delta)$ (satisfying \eqref{equ:BW}) and estimate the stage 2 parameters $\pmb{\beta}_2$ and $\pmb{\Psi}_2$ by solving the following GEEs
    \begin{eqnarray*}
    U_2(\pmb{\Psi}_2, \pmb{\beta}_2) &=& \sum_{i=1}^{n} \delta_i \eta_{i2} \hat{w}_{i2} \begin{pmatrix} \mathbf{h}_{i2}^{\pmb{\beta}} \\ a_{i2} \mathbf{h}_{i2}^{\pmb{\Psi}} \end{pmatrix}\\
    && \quad \boldsymbol{\cdot} \left( \log(Y_{i2}) - \pmb{\beta}_2^\top \mathbf{h}_{i2}^{\pmb{\beta}} - a_{i2} \pmb{\Psi}_2^\top \mathbf{h}_{i2}^{\pmb{\Psi}} \right)\\
    &=& 0.
    \end{eqnarray*}
    
    \item Given the estimated parameters and the resulting optimal decision in Step 1, construct the stage 1 pseudo-outcome according to \eqref{equ:psudo}. Then, specify weights $w_1(\mathbf{h}_1, a_1, \delta)$ and estimate the stage 1 parameters $\pmb{\beta}_1$ and $\pmb{\Psi}_1$ by solving the following weighted GEE
\begin{eqnarray*}
&& U_1(\pmb{\Psi}_1, \pmb{\beta}_1; \hat{\pmb{\Psi}}_2) = \sum_{i=1}^{n} \delta_i \hat{w}_{i1} \begin{pmatrix} \mathbf{h}_{i1}^{\pmb{\beta}} \\ a_{i1} \mathbf{h}_{i1}^{\pmb{\Psi}} \end{pmatrix}\\
&& \qquad \qquad \boldsymbol{\cdot}  \left( \log(\tilde{Y}^{a_1,a_2^{opt}}) - \pmb{\beta}_1^\top \mathbf{h}_{i1}^{\pmb{\beta}} - a_{i1} \pmb{\Psi}_1^\top \mathbf{h}_{i1}^{\pmb{\Psi}} \right) \\
&& \qquad \qquad \qquad \, \, = 0.
\end{eqnarray*}
\end{enumerate}

Simoneau et al. \cite{simoneau2020optimal} developed the \texttt{DWSurv} function within the DTRreg package to simplify the implementation of the DWSurv method. 

\subsection*{Applications}

 Simoneau et al. \cite{simoneau2020estimating} illustrated the DWSurv approach using a dataset on rheumatoid arthritis (RA) patients. RA is a chronic condition characterized by recurrent episodes of high disease activity followed by periods of remission, necessitating adaptive treatment strategies. At each stage of treatment, patients can either continue with traditional disease-modifying antirheumatic drug (DMARD) monotherapy or switch to a DMARD combination therapy, with the main outcome of interest being the time to remission at the time patients enter a period of disease flare-up. The primary goal of DTR estimation in this example is to identify optimal sequences of treatment decisions that minimize the time to remission for individuals based on their evolving clinical characteristics.

In the study by Zhang et al. \cite{zhang2020individualized}, fluid management for critically ill sepsis patients was structured as a DTR problem applied over three stages: Day 1, Day 3, and Day 5 post-ICU admission. The primary outcome was the time from ICU admission to discharge or death. At each stage, the fluid treatment was categorized as either liberal ($\geq$ 40 ml/kg/day) or restricted ($<$ 40 ml/kg/day), with the choice based on patient characteristics. Results demonstrated that adapting fluid intake strategies to patient-specific information via DWSurv significantly prolonged survival compared to actual treatment approaches.

Simoneau et al. \cite{simoneau2020adaptive} applied the DWSurv approach to estimate optimal DTRs for managing type 2 diabetes. The study focused on sequentially optimizing drug add-ons for patients with inadequate glycemic control on metformin monotherapy.  The treatment involved adding sulfonylurea or another drug initially, then adjusting to dipeptidyl peptidase-4 inhibitors or others, with the primary outcome being the time to a cardiovascular event or death. The results of this study indicate that using DWSurv to tailor treatment decisions based on patient characteristics, including glycemic control and body mass index (BMI), could potentially improve long-term outcomes in type 2 diabetes management.

Other applications of the DWSurv approach in DTR estimation problems can be found in Coulombe et al. \cite{coulombe2021can} and Moodie et al. \cite{moodie2022privacy} with a focus on maximizing the time to severe depression in antidepressant therapies.

 \subsection{DWSurvMT: dWOLS with survival outcomes and multicategory treatments}

 \label{sec:DWSurvMT}
   So far, dWOLS-based methods have focused on binary treatment decisions or continuous dose-finding scenarios.  However, there are scenarios where patients are prescribed one of multiple treatment options at each stage. Zhang et al. \cite{zhang2022doubly} extended the original dWOLS to handle multicategory treatment options at various stages while considering the survival time as the primary outcome (similar to DWSurv). Unlike the other dWOLS-based methods, which define the blip function as the difference in expected outcomes between treatment and control arms, Zhang et al. \cite{zhang2022doubly} redefined it to compare the expected outcome for a given treatment to a weighted average across all treatments. This approach eliminates the need to select a "reference" treatment, which can be challenging in multicategory treatment scenarios. 
  
\subsection*{Notations and blip model}

Zhang et al. \cite{zhang2022doubly} proposed defining the blip function for any treatment $a_j$  as follows
\begin{eqnarray}\label{equ:BLIP}
 \gamma_j(\mathbf{h}_j, a_j; \pmb{\Psi}_{j a_j}) &=& Q_j(\mathbf{h}_j, a_j) \\
&& - \sum_{a \in \mathcal{A}_j} m_j(a) Q_j(\mathbf{h}_j, a).\nonumber
\end{eqnarray}

In equation \eqref{equ:BLIP},   $Q_j(\mathbf{h}_j, a_j)$ (with $Q$ representing the quality function) is the expected outcome model given in \eqref{equ:EOM}, and $m_j(a) > 0$ is a prespecified weight function for treatment $a_j$ in the treatment space $\mathcal{A}_j=\{1,\ldots, N_j\}$ satisfying $\sum_{a \in \mathcal{A}_j} m_j(a) = 1$, where $N_j$ is the number of available treatment options at stage $j$. The blip function in \eqref{equ:BLIP} represents the difference between the value of the Q-function for treatment $a_j$ and the weighted average of Q-functions over all treatments. Based on this definition, the optimal treatment decision at stage $j$ is determined solely by the blip function in \eqref{equ:BLIP}, such that $a_j^{\text{opt}} = d^*_j(\mathbf{h}_j) = \underset{a_j \in \mathcal{A}_j}{\arg\max} \, \gamma_j(\mathbf{h}_j^{\Psi}, a_j; \pmb{\Psi}_{j a_j})$.

This formulation allows the expected outcome model to be written as
\begin{align*}
Q_j(\mathbf{h}_j, a_j) = \sum_{a \in \mathcal{A}_j} m_j(a) Q_j(\mathbf{h}_j, a) + \gamma_j(\mathbf{h}_j^{\Psi}, a_j; \pmb{\Psi}_{j a_j}),
\end{align*}
where the first term on the right-hand side is the weighted average over all treatments and can be considered the treatment-free component. Consistent with the existing dWOLS methodology, both the treatment-free component and the blip function can be modeled linearly as
\begin{eqnarray}\label{equ:newBLIP}
&&\sum_{a \in \mathcal{A}_j} m_j(a) Q_j(\mathbf{h}_j, a) = \boldsymbol{\beta}_j^\top \mathbf{h}_j^{\boldsymbol{\beta}}\nonumber\\ &&\text{and} \quad  \\
&&\gamma_j(\mathbf{h}_j^{\Psi}, a_j; \pmb{\Psi}_{j a_j}) = \boldsymbol{\Psi}_{j a_j}^\top  \mathbf{h}_j^{\boldsymbol{\Psi}}\nonumber.
\end{eqnarray}

In this representation, it is clear that the treatment-free component is independent of the assigned treatment $a_j$ and is shared across the treatment space $\mathcal{A}_j$. The main distinction between the blip parameterization in \eqref{equ:newBLIP} and those discussed earlier is that, in the earlier models, the treatment decision $a_j$ directly influenced the blip, and the parameter $\pmb{\Psi}_j$ was fixed across different treatments. However, the blip parameterization in \eqref{equ:newBLIP} allows the parameters to vary with treatments, which is why $\pmb{\Psi}_j$ is indexed by $a_j$ in addition to the stage index $j$. This means that the treatment-specific coefficients $\pmb{\Psi}_{j a_j}$, which are also called targeted blip parameters by Zhang et al. \cite{zhang2022doubly}, depend on both $\mathbf{h}_j^{\pmb{\Psi}}$ and $a_j$.

Given that $\sum_{a \in \mathcal{A}_j} m_j(a) = 1$, Zhang et al. \cite{zhang2022doubly} demonstrated that $\sum_{a \in \mathcal{A}_j} m_j(a) \boldsymbol{\Psi}_{j a} = 0$. Thus, incorporating this finding and equation \eqref{equ:newBLIP}, the Q-functions (expected outcome model) can be represented as
\begin{align}\label{equ:quality}
Q_j(\mathbf{h}_j, a_j) = \pmb{\beta}_j^\top \mathbf{h}_j^{\pmb{\beta}} + \pmb{\Psi}_{j a_j}^\top \mathbf{h}_j^{\pmb{\Psi}},
\end{align}
with the constraint $\sum_{a \in \mathcal{A}_j} m_j(a) \pmb{\Psi}_{j a} = 0$. They also show that the the quality function in \eqref{equ:quality} can be represented as
\begin{align}\label{equ:blip5}
Q_j(\mathbf{h}_j, a_j)= \pmb{\beta}_j^\top \mathbf{h}_j^{\pmb{\beta}} + \sum_{a \neq N_j} \pmb{\Psi}_{j a}^\top \tilde{\mathbf{h}}_{j a}^{\pmb{\Psi}}, 
\end{align}
with $\tilde{\mathbf{h}}_{j a}^{\pmb{\Psi}} = \left( I(a_j = a) - I(a_j = N_j) m_j(N_j)^{-1} m_j(a) \right) \mathbf{h}_j^{\pmb{\Psi}}$. Therefore, equation \eqref{equ:blip5} can be viewed as a simple linear regression with covariates $\{\mathbf{h}_j^{\pmb{\beta}}, \tilde{\mathbf{h}}_{j1}^{\pmb{\Psi}}, \ldots, \tilde{\mathbf{h}}_{j(N_j-1)}^{\pmb{\Psi}}\}$.

Finally, recall that the observed data under this approach, which follows the same structure as in DWSurv, are of the form $(\eta_{i1}, x_{i1}, a_{i1}, y_{i1}, \ldots, \eta_{iK}, x_{iK}, a_{iK}, y_{iK}, \delta_i)$, where $\eta_{ij}$ is an indicator representing whether individual $i = 1, \ldots, n$ entered stage $j = 1, \ldots, K\), with \(\eta_{i1} = 1$ for all $i$, and $\delta_i$ is the censorship indicator.  

 \subsection*{Balancing weights}
 
 To ensure double robustness, Zhang et al. \cite{zhang2022doubly} demonstrated that solving weighted GEEs produces doubly robust estimates of the targeted blip functions. This holds true when the balancing weight $w(\mathbf{h}, a, \delta)$ satisfy the following condition
\begin{eqnarray}\label{equ:balancweight}
&&\frac{g(\mathbf{h}, a) \pi(\mathbf{h}, a; \pmb{\alpha}) w(\mathbf{h}, a, 1)}{m(a)} =\\
&& \qquad \qquad \qquad \quad \frac{g(\mathbf{h}, a') \pi(\mathbf{h}, a'; \pmb{\alpha}) w(\mathbf{h}, a', 1)}{m(a')}\nonumber,
\end{eqnarray}
for any $a, a' \in \mathcal{A}$   where $\pi(\mathbf{h}, a; \pmb{\alpha})=\mathbb{P}(A=a|\mathbf{h}, \eta=1; \pmb{\alpha})$ is the treatment model. It is also worth mentioning that different choices of weights $\{m(a): a \in \mathcal{A}\}$ lead to different targeted blip parameters; however, they do not affect the theoretically optimal treatment rules, which are the main focus of DTR estimation. Zhang et al. \cite{zhang2022doubly} discussed two options for these weights: $m(a) = N^{-1}$ and $m(a) = \pi(a)$ for any $a \in \mathcal{A}$. It will be shown later that the weight $m(a)$ is primarily used to determine the balancing weights in \eqref{equ:balancweight} required to implement the weighted least squares algorithm for parameter estimation. Once the balancing weights are identified, there is no need to specify the treatment-specific weights $m(a)$ in the DWSurvMT implementation.

\subsection*{Implementation procedure}

 Zhang et al. \cite{zhang2022doubly} proposed the following steps to estimate the targeted blip parameters using the dWGLM approach.

\begin{itemize}
    \item[Step 1] (Estimation at the final stage $j = K$)  Estimate the stage $K$ blip parameters $\pmb{\Psi}_{K1}, \ldots, \pmb{\Psi}_{KN_j}$ by solving the following GEEs
    \begin{eqnarray*}
    &&\sum_{i \colon \eta_{iK} = 1} \delta_i \hat{w}_{iK} \Big\{ \log(y_{iK}) - \pmb{\beta}_K^\top \mathbf{h}_{iK}^{\pmb{\beta}}\\
    && \qquad - \sum_{a \neq N_K} \pmb{\Psi}_{K a}^\top \tilde{\mathbf{h}}_{iK a}^{\pmb{\Psi}} \Big\} \mathbf{h}_{iK}^{\pmb{\beta}} = 0,\\
    && \sum_{i \colon \eta_{iJ} = 1} \delta_i \hat{w}_{iK} \Big\{ \log(y_{iK}) - \pmb{\beta}_J^\top \mathbf{h}_{iK}^{\pmb{\beta}}\\
    &&\qquad - \sum_{a \neq N_K} \pmb{\Psi}_{K a}^\top \tilde{\mathbf{h}}_{iK a}^{\pmb{\Psi}} \Big\} \tilde{\mathbf{h}}_{iK a}^{\pmb{\Psi}} = 0,
    \end{eqnarray*}
    for $a = 1, \ldots, N_K - 1$.

 \item[Step 2] (Recursive estimation)  For $j = K - 1, \ldots, 1$, construct the stage $j$ pseudo-outcome as
        \begin{eqnarray*}
        \tilde{y}_{ij} &=& y_{ij} + \tilde{y}_{i(j+1)} \exp \Big( - \sum_{a \neq N_{j+1}} \hat{\pmb{\Psi}}_{(j+1) a}^\top \tilde{\mathbf{h}}_{i(j+1) a}^{(1)} \\
        && + \max_{a_{j+1} \in \mathcal{A}_{j+1}} \sum_{a \neq N_{j+1}} \hat{\pmb{\Psi}}_{(j+1) a}^\top \tilde{\mathbf{h}}_{i(j+1) a}^{(1)} \Big),
        \end{eqnarray*}
        with \(\tilde{y}_{iK} = y_{iK}\). Then, estimate the stage $j$ blip parameters $\pmb{\Psi}_{j1}, \ldots, \pmb{\Psi}_{jN_j}$  by the linear regression of $\log(\tilde{y}_{ij})$, that is, solving the following GEEs
        \begin{eqnarray*}
        && \sum_{i \colon \eta_{ij} = 1} \delta_i \hat{w}_{ij} \Big\{ \log(\tilde{y}_{ij})\\
        && \qquad - \pmb{\beta}_j^\top \mathbf{h}_{ij}^{\pmb{\beta}} - \sum_{a \neq N_j} \pmb{\Psi}_{j a}^\top \tilde{\mathbf{h}}_{ij a}^{\pmb{\Psi}} \Big\} \mathbf{h}_{ij}^{\pmb{\beta}} = 0,\\
        && \sum_{i \colon \eta_{ij} = 1} \delta_i \hat{w}_{ij} \Big\{ \log(\tilde{y}_{ij})\\
        && \qquad - \pmb{\beta}_j^\top \mathbf{h}_{ij}^{\pmb{\beta}} - \sum_{a \neq N_j} \pmb{\Psi}_{j a}^\top \tilde{\mathbf{h}}_{ij a}^{\pmb{\Psi}} \Big\} \tilde{\mathbf{h}}_{ij a}^{(1)} = 0, 
        \end{eqnarray*}
        for  $a = 1, \ldots, N_j - 1$.
 \end{itemize}

\subsection*{Applications}

Zhang et al. \cite{zhang2022doubly} illustrated the proposed method by applying it to real-world data from the Standard and New Antiepileptic Drugs (SANAD) study of patients with epilepsy. The primary outcome of interest was survival time.  At each decision point, patients were assigned one of the following three available treatment options: lamotrigine, carbamazepine, and a third unspecified drug. The authors concluded that the proposed method led to more accurate predictions of the optimal treatment regime when compared to existing methods. They also found that patients who followed the optimal DTRs estimated by their approach showed improved survival outcomes, particularly favouring lamotrigine over the other options.

 \section{Assesing and extending the dWOLS assumptions }
 \label{sec:assess}

This section reviews studies that evaluate key dWOLS assumptions, examine scenarios where violations of these assumptions occur, and outline reasonable approaches to address these issues. Notably, all the studies reviewed here assume a continuous outcome and a binary treatment decision at each stage.

 \subsection{dWOLS and model validation and variable selection}
The effectiveness of the dWOLS approach to finding optimal DTRs relies on the correct specification of the models involved—namely, the treatment model, the treatment-free model, and the blip function. The inaccurate model specification can lead to biased estimates of the treatment effects model (blip), compromising the validity of the derived optimal treatment rules. Given this dependency, model validation and selection become crucial components of the dWOLS methodology to ensure robust and reliable conclusions. Wallace et al. \cite{wallace2017model} focused on model validation and selection within the context of DTR estimation using the dWOLS approach.

Model validation involves determining whether a given model is correctly specified by evaluating the consistency and stability of parameter estimates across different model specifications. It is important to note that the treatment model is generally easier to assess using standard model-checking techniques, such as those applied in logistic regression. In contrast, the treatment-free model is more challenging to evaluate. Wallace et al. \cite{wallace2017model} addressed this difficulty by proposing an innovative model assessment approach that leverages the double robustness property of the dWOLS method. The assessment strategy involves running dWOLS with different treatment models while keeping the treatment-free model the same. If the treatment-free model is correctly specified, all analyses should yield similar (consistent) blip parameter estimates. However, if the treatment-free model is misspecified, the blip estimates may differ significantly between the analyses. By comparing these estimates, analysts can gain valuable insights into the validity of the models. This approach offers a practical way to detect misspecification in either the treatment model or the treatment-free model, without relying solely on standard model-checking techniques.

In contrast, model selection is the process of choosing the best model among a set of candidate models. The authors propose the use of the quasi-likelihood information criterion (QIC) as a tool for model selection, arguing that traditional likelihood-based criteria, such as the Akaike information criterion (AIC), may not be appropriate due to the non-likelihood-based nature of dWOLS. QIC, analogous to AIC, is tailored to the dWOLS framework and helps identify the ``best" model by minimizing the Kullback-Leibler divergence between a proposed model and the true model. This approach provides a formal way to select models based on their performance, ensuring the correct specification of both the treatment and treatment-free models for dynamic treatment regime estimation.

Including irrelevant covariates in the treatment model reduces statistical efficiency and complicates clinical interpretation of treatment rules. Variable selection is therefore essential for identifying only the relevant tailoring variables. While dWOLS does not mandate identical covariates in both the blip and treatment-free components, any covariate interacting with the treatment in the blip must also appear in the treatment-free component (the strong heredity principle). Bian et al. \cite{bian2023variable} introduced penalized dWOLS (pdWOLS) as an extension of dWOLS to overcome limitations like the lack of adherence to the strong heredity principle. pdWOLS incorporates penalty terms into the estimation process, facilitating simultaneous parameter estimation and variable selection to maintain the strong heredity assumption. The authors also establish the double robustness of the pdWOLS estimators. By doing so, pdWOLS offers a more refined methodology that not only improves computational efficiency but also enhances clinical applicability.

 \subsection{dWOLS in the presence of error-prone covariates}
 
The original dWOLS assumes that all variables—main outcome, treatments, and covariates—included in the DTR estimation are measured without any errors. However, this assumption may not hold for some variables in real-world scenarios, as true values often go unobserved due to various error sources. For example, self-reported health data are prone to errors. Analyzing data without correcting for these errors can produce unreliable results. Specifically, in DTR estimation, ignoring measurement errors can bias the blip model estimates, leading to suboptimal treatment recommendations.

Spicker and Wallace \cite{spicker2020measurement} investigated DTR estimation using the dWOLS method in the presence of measurement errors in some covariates. They assumed that some other covariates, along with the main outcome and treatment variables, are measured without error. Consider a simple scenario involving two covariates: an error-prone covariate $X^*$ modeled using the classical additive error model $X^* = X + U$  where $X$ represents the (unobserved) true covariate and $U$ is an error term and an error-free covariate $Z$, both influencing the outcome $Y$. In this setting, the main outcome $(Y)$ is affected by the true value of the covariate ($X$) through a biological process, while the treatment decision is based on the observed (error-prone) covariate, i.e., $a=d(X^*)$. To account for the effect of error-prone covariates,  Spicker and Wallace \cite{spicker2020measurement} used the regression calibration approach in which $X$ is imputed using an estimator $\widehat{X} \quad (e.g., \widehat{X}=\mathbb{E}(X|Z, X^*))$, which is a function of the observed information $X^*$ and $Z$. Then, the standard dWOLS can be applied to estimate the optimal DTR using $\widehat{X}$s.

 Regression calibration has been shown to produce consistent estimators in linear models. However, in non-linear models, this approach does not generally guarantee consistency for all parameters but can effectively reduce bias in some of them.  For example, in our treatment model estimation where the primary goal is to estimate the probabilities rather than the coefficients, regression calibration offers a reasonable approximation of the probabilities. We know that to apply dWOLS, we must specify the balancing weights.  In the error-prone setting, if we employ regression calibration using $\widehat{X}$ in our outcome models, we wish to induce covariate balance, not between $X$ and $A$, but between $\widehat{X}$ and $A$. Accordingly, we might speculate that any weights which satisfy  $\pi(\widehat{X}; \pmb{\alpha})w(\widehat{X}, 1)=(1-\pi(\widehat{X},\pmb{\alpha}))w(\widehat{X}, 0)$ will induce covariate balance in $\widehat{X}$.

 While previous discussions addressed error correction for estimating DTRs, the challenge remains for treatment decisions in future patients with error-prone covariates not included in the estimation data. Some literature suggests using error-prone covariates as predictors for treatment decisions, such as averaging measurements ($X = \frac{1}{k} \sum_{l=1}^k X^*_l$) as proxies. Spicker and Wallace \cite{spicker2020measurement} explored this approach, but prior research suggests that ignoring measurement errors can still yield biased results. These complexities in the DTR prediction context highlight the need for further research on error correction techniques.
 
 \subsection{dWOLS in the presence of interference}
 
Interference in the context of DTRs occurs when the outcome of an individual is influenced not only by their own covariates and treatments but also by the treatments assigned to other individuals, often their network neighbours. This violates the SUTVA, a key assumption in DTR estimation that assumes the outcome for one individual is unaffected by others' treatments. However, in real-world social network contexts—such as family units, peer groups, or public health interventions—this assumption is often violated. When interference is present, standard DTR estimation methods like dWOLS, which relies on SUTVA, may yield biased estimators leading to suboptimal treatment decisions, as the treatment effects could spill over from neighbours to individuals. 

Jiang et al. \cite{jiang2023dynamic} extended the original dWOLS approach to account for possible interference. The motivation for their work comes from the PATH study, where the goal was to examine smoking cessation behaviours within two-person household networks. In this dataset, the treatment (participation in a smoking cessation program) of one household member may influence the other, violating the no-interference assumption. The paper introduces the concept of a \textit{network propensity function} to address this issue. This function estimates the probability of a treatment assignment within a network structure, capturing the spillover (indirect) effects from neighbours' treatments. The authors demonstrate that by incorporating these network propensity functions into the dWOLS framework, double robustness is preserved.

One key innovation of this paper is the introduction of \textit{network weights}, which adjusts for the spillover effects of treatments across neighbours in a network. These network weights balance the covariates of individuals and their neighbours, ensuring that the treatment assignment mechanism is well-specified in the presence of interference. The authors provide theoretical guarantees for the consistency of their estimators under interference and illustrate the effectiveness of their approach through both simulations and real-world data from the PATH study.

In the analysis of the PATH dataset, the authors examined smoking cessation behaviours and found that standard DTR methods failed to account for the influence of one household member's smoking behaviour on the other. By applying their extended dWOLS method with network propensity scores, they demonstrated improved accuracy in treatment effect estimation. Interference in this case came from the direct influence one household member had on the other’s smoking cessation success. In this study, the primary outcome of interest was cigarette dependence, measured as a continuous score, and the treatment was binary (program participation or not), observed over multiple decision points.

 \subsection{dWOLS in the presence of non-regularity}
Non-regularity in DTR estimation refers to the violation of standard regularity conditions that ensure smooth and well-behaved asymptotic distributions for estimators, which typically exhibit asymptotic normality. In non-regular settings, however, these assumptions break down, leading to estimators with non-normal limiting distributions. This non-regularity poses significant challenges for constructing reliable confidence intervals and performing inference, as standard methods such as Wald-type confidence intervals or bootstrap methods may fail to provide accurate coverage rates. One major source of non-regularity in DTR estimation, as Simoneau et al. \cite{simoneau2018non} identified, occurs when treatment effects at each stage are small or near zero. When the stage-specific treatment effects are weak, the differences in expected outcomes between treatment options are minimal, resulting in ambiguous or non-unique treatment decisions. For instance, when blip parameters approach zero, both treatment choices may seem equally optimal, creating non-smooth maximization problems that propagate non-regularity through the estimation process.

In the context of DTR estimation using dWOLS, non-regularity arises from the non-smooth nature of the decision-making process at each stage of treatment. To illustrate, consider a two-stage treatment regime. As mentioned earlier,  the pseudo-outcome at the first stage depends on the second-stage blip parameter $\pmb{\Psi}_2$. The function that determines the optimal treatment at the second stage, $I(\hat{\pmb{\Psi}}_2^\top \mathbf{h}_2^{\pmb{\Psi}} > 0)$, is non-differentiable at the point where $\hat{\pmb{\Psi}}_2^\top  \mathbf{h}_2^{\pmb{\Psi}} = 0$. This non-smoothness at the point of non-differentiability leads to problems in the estimation process because the estimate of the first-stage blip parameter $\pmb{\Psi}_1\) is obtained through a weighted OLS regression on a pseudo-outcome $\tilde{y}_1$ that is itself a non-smooth function of $\hat{\pmb{\Psi}}_2$. As a result, the DTR estimator of $\pmb{\Psi}_1$ is non-regular, and thus the asymptotic distribution of the pivotal quantity $\sqrt{n} (\hat{\pmb{\Psi}}_1 - \pmb{\Psi}_1)\) is not expected to behave uniformly normal.

The non-regularity in this context depends on how close the second-stage blip parameter $\pmb{\Psi}_2$ is to the point of non-differentiability. If the probability of generating a history $ \mathbf{h}_2^{\pmb{\Psi}}$ such that $\pmb{\Psi}_2^\top  \mathbf{h}_2^{\pmb{\Psi}} = 0$ is zero, the asymptotic distribution of the first-stage estimator is normal. However, if this probability is non-zero, meaning that the data frequently generates scenarios where the second-stage treatment decision is unclear (i.e., the two possible treatments yield nearly identical expected outcomes), the asymptotic distribution becomes non-normal. This probability, denoted by $p= \mathbb{P}(\mathbf{H}_2^{\pmb{\Psi}} : \pmb{\Psi}_2^\top  \mathbf{h}_2^{\pmb{\Psi}} = 0)$, serves as a measure of non-regularity in the data. If $p > 0$, the estimation process exhibits non-regularity, and standard inference techniques may fail.

This study was motivated by a dataset in which small treatment effects at certain stages introduced non-regularity into the estimation of treatment parameters. The authors investigated the impact of infant diet on long-term health outcomes, focusing specifically on the timing of introducing solid food during the first year of life. The primary outcomes of interest were the child’s BMI, waist circumference, and triceps skinfold thickness measured at 6.5 years of age (all continuous variables). The treatment options available at each decision point were the introduction of solid food between $3-6$ months and $6-9$ months. This analysis used data from the PROmotion of Breastfeeding Intervention Trial (PROBIT), which included longitudinal measurements from over 17,000 infant-mother pairs. In this example, non-regularity likely arises due to the small effect of solid food intake between $3-6$ months and $6-9$ months on the outcomes measured at age $6.5$. As a result, inferences using dWOLS may yield estimators with non-regular limiting distributions, leading to poorly performing standard confidence intervals.

To address the non-regularity issue in practice, Simoneau et al. \cite{simoneau2018non} proposed using the $m$-out-of-$n$ bootstrap method, which is specifically designed to handle non-regular situations. The $m$-out-of-$n$ bootstrap involves resampling a subset of the original data (of size $m < n$) and constructing confidence intervals based on these smaller resamples. This method is more robust in non-regular settings because it does not rely on the smoothness of the estimator, unlike standard bootstrap methods. By choosing an appropriate resample size $m$, which reflects the degree of non-regularity in the data, the $m$-out-of-$n$ bootstrap can provide more accurate confidence intervals. In particular, the authors suggested an adaptive approach for selecting $m$, based on the probability $p$ of generating a history where $\pmb{\Psi}_2^{T} \mathbf{h}_2^{\pmb{\Psi}}  = 0$. This approach ensures that the resample size is adjusted according to the degree of non-regularity.

\section{Numerical Examples}
\label{sec:NE}

This section presents hypothetical numerical examples using simulated data, along with step-by-step implementations of various dWOLS-based methods in the \texttt{R} environment. These examples aim to make the discussed methodologies more intuitive and provide practical guidance for implementing the methods across diverse DTR estimation problems.

We demonstrate the original dWOLS method discussed in Section \ref{sec:dWOLS} and its four extensions—G-dWOLS, dWGLM, DWSurv, and DWSurvMT—described in Section \ref{sec:Ext}. Each example assumes a two-stage treatment regime $(K=2)$ with one covariate measured at each stage for every individual and a population of size $n=10,000$ individuals. The code for implementing the dWOLS method is provided in Algorithm 1, while detailed implementations for the other methods can be found in Appendices A–D (Algorithms 2–5). To facilitate better understanding, we have included explanatory notes within the code to describe the purpose and functionality of each line. Additionally, some functions have been omitted from some algorithms to save space. However, we provide the complete code for all five methods as supplementary files to ensure readers can easily reproduce the results discussed here.

\begin{figure*}
\includegraphics[scale=0.63]{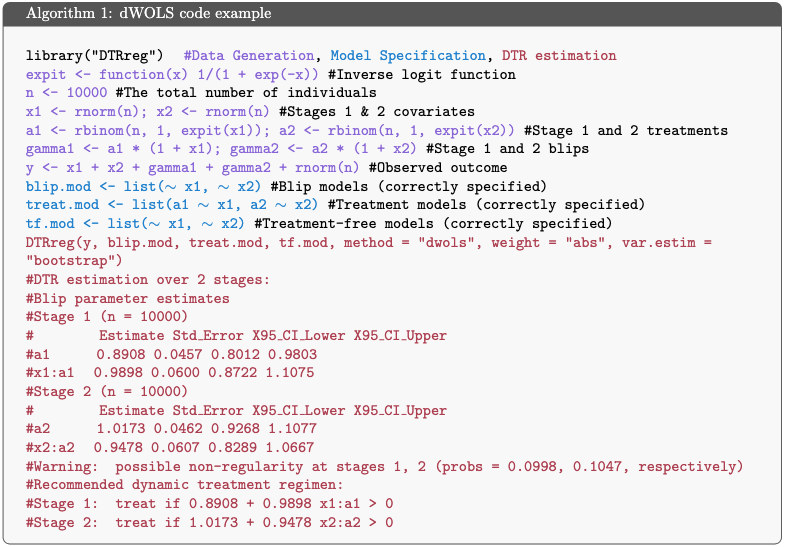}
\label{penG}
\end{figure*}

We further explore various scenarios, including linear and nonlinear models for the treatment-free models, as well as cases with model misspecifications. Initially, we consider the simplest dWOLS setting, assuming linear forms for both the treatment-free and blip models. Specifically, we set the blip parameters as $\Psi_{10} = \Psi_{11} = \Psi_{20} = \Psi_{21} = 1$, the treatment-free parameters as $\beta_{10} = \beta_{11} = \beta_{20} = \beta_{21} = 1$, and the treatment model parameters as $\alpha_{10} =  \alpha_{20} =0$ and $\alpha_{11} = \alpha_{21} = 1$. The balancing weights are also assumed to be the absolute difference $w(x, a)=|a-E(A|x)|$.  We summarize this setting as follows
\begin{itemize}
    \item The treatment-free model at stage $j=1,2$ is
    \begin{align*}
        f_j\left(x_j; \pmb{\beta}_j=(1,1)\right)= 1 + x_j.
    \end{align*}
    \item The blip model at stage $j=1,2$ is $\gamma_j(x_j, a_j; \pmb{\Psi}_j=(1,1))=  a_j (1 + x_j)$.
    \item Treatment model at stage $j=1,2$ is $\pi_j(x_j;\pmb{\alpha}_j)=\mathbb{P}(A_j=1|x_j; \pmb{\alpha}_j=(0,1))=(1+e^{- x_j})^{-1}$.
\end{itemize}

Algorithm 1 provides the dWOLS code example. Note that in this example, all the models are correctly specified. The blip estimates are displayed for two stages at the end of the output along with 95\% confidence intervals shown on the right-hand side for their respective parameters. The output in the algorithm largely speaks for itself. According to the resulsts, the estimated blip parameters are $(\widehat{\Psi}_{10}, \widehat{\Psi}_{11}) = (0.8908, 0.9898)$ for stage 1 and $(\widehat{\Psi}_{20}, \widehat{\Psi}_{22}) = (1.0173, 0.9478)$ for stage 2. These estimates guide the recommended treatment regimen summarized at the end of the output. For instance, for a patient with stage 1 and 2 covariates $x_1=1$ and $x_2=-1.5$, the estimated DTR prescribes treatment ($a_1=1$) at stage one and no treatment ($a_2=0$) at stage two.

\section{Research gap and future works}

The dynamic weighted ordinary least squares method was introduced as an approach to estimating dynamic treatment regimes. This method is distinguished by its simplicity of implementation and robustness against model misspecification.  A decade after its initial development, this paper offers a comprehensive review of advancements in dWOLS-based methods for DTR estimation, spanning both theoretical and practical aspects. To provide a clear overview of the research landscape of theoretical developments on this topic, Table \ref{tab:LR} summarizes and categorizes the relevant papers based on the outcome and treatment types, the diseases analyzed (illustration), and the primary objective of each study. This table serves as a tool for identifying current gaps in the literature and highlighting potential avenues for future research.

 \begin{sidewaystable}
\scriptsize
\setlength{\tabcolsep}{1.5pt}
\renewcommand{\arraystretch}{1.5}
\caption{Summary of key studies on the dWOLS approach and the related advancements.}
\begin{tabular}{lcccccccccccc}
\hline\hline
Paper & Abbreviation & Outcome type & Treatment type &  Illustration & Primary objective  & &  &  & &  &   &   \\
\hline
Wallace and Moodie \cite{wallace2015doubly} & dWOLS & Continous  &Binary &  Breastfeeding interventions &  DTR estimation & & &  & &  && \\
Wallace et al. \cite{wallace2016package, wallace2017dynamic} & dWOLS & Continous  & Binary &  ---  & introducing the R package \texttt{DTRreg} for dWOLS application & & & & &  && \\
 Wallace et al. \cite{wallace2017model} & dWOLS & Continous  & Binary &  Depression relief  &  Model selection and validation in dWOLS & & & & &  && \\
Simoneau et al. \cite{simoneau2018non} & dWOLS & Continous  & Binary &  Breastfeeding interventions  &  DTR estimation with non-regular inference & & & & &  && \\
Simoneau et al. \cite{simoneau2020estimating} &DWSurv & Time-to-event (survival) with censorship  & Binary  &  Rheumatoid arthritis &  DTR estimation  &  & & & &  && \\
Simoneau et al. \cite{simoneau2020optimal} &DWSurv & Time-to-event (survival) with censorship  & Binary  &  --- &  Introducing \texttt{DWSurv} function in the \texttt{DTRreg} for DWSurv application  &  & & & &  && \\
 Spicker and Wallace \cite{spicker2020measurement} & DWOLS& Continous  & Binary  & Depression relief   & DTR estimation with error-prone covariates &     &  & & &  && \\
 Schulz and Moodie \cite{schulz2021doubly} & G-dWOLS  & Continous  & Continous &  Thrombosis prevention &   DTR estimation & & & & &  && \\
Zhang et al. \cite{zhang2022doubly}  & DWSurvMT & Time-to-event (survival) with censorship  &  Multicategory   &  Epilepsy &  DTR estimation   &   & & & & &  & \\
Jiang et al. \cite{jiang2022doubly}  & DWGLM & Binary  & Binary &  Smoking cessation &  DTR estimation  & & & & &  && \\
Jiang et al. \cite{jiang2023dynamic}  & dWOLS & Continous  & Binary & Smoking cessation & DTR estimation with interference &   & & & & &  & \\
Bian et al. \cite{bian2023variable}  & pdWOLS & Continous   & Binary & Depression relief  & Variable selection and DTR estimation  &  & & & & &  & \\
\hline\hline
\end{tabular}\label{tab:LR}
\end{sidewaystable}

As this table illustrates, there is of course much room for the literature around the dWOLS method to expand. For example, much of the methodological development to date has focused on binary treatments, albeit with notable exceptions highlighted in Sections \ref{sec:Ext} and \ref{sec:assess}. Nevertheless, expanding dWOLS to account for non-binary treatments when the outcome is binary would seem like a particularly important avenue of exploration, especially given the heavy focus on binary outcomes within health settings. Additionally, there are diseases with outcome/treatment combinations that fall outside the scope of current dWOLS-based methods. Addressing these gaps by adapting dWOLS to handle such cases represents another important direction for future research. For instance, investigating the dWOLS for continuous outcomes and categorical treatments.

It should also be noted that while many of the expansions concerning model limitations (such as measurement error and model checking) are focused on binary treatments also, there are clear opportunities to take these existing methods and expand them into more varied variate types. For example,  one valuable area for exploration could be model selection and validation in dosage-finding problems (continuous treatments) as the blip models in these scenarios can adopt more complex forms compared to those used for binary treatments. Another critical area for further research is a comprehensive investigation of the effects of measurement errors on the dWOLS-based methods. Currently, only one study has examined the impact of measurement error in covariates on the original dWOLS method performance. However, outcomes and treatments themselves can also be error-prone, especially when measured using devices. Such devices often produce inaccurate or imprecise readings, which can result in unreliable estimates of the DTR. Consequently, investigating the effects of measurement errors in outcomes and/or treatments on the dWOLS-based method, along with developing strategies to mitigate these errors' impact, constitutes an interesting area for future research. Given the developments seen over the past decade, we expect to see these, and other, future directions for dWOLS pursued in the coming years.

\begin{appendix}
\section*{Appendix A: G-dWOLS code example}\label{appn1}

We provide here the example code for the G-dWOLS approach discussed in Subsection \ref{sec:G-dWOLS}, demonstrating its application in a scenario where the treatment-free models are assumed to be non-linear, and the blip functions are quadratic with respect to $a$. Additionally, the treatment dosage $a$ is modeled as following a normal distribution, with its mean varying based on the covariate $x$. The balancing weights are assumed to be the inverse probability of treatment. This setting is summarized as follows
  \begin{itemize}
    \item The treatment-free model at stage $j=1,2$ is $f_j(x_j)=log(x_j)+sin(x_j)$. 
    \item The blip model at stage $j=1,2$ is $\gamma_j(x_j, a_j; \pmb{\Psi}_j=(1, 1, -1))= a_j(1+ x_{j}) - a_j^2 $.
    \item Treatment model at stage $j=1,2$ is  $A_j \sim N(\alpha_{j0}+\alpha_{j1} X_j, 1)$ where $\alpha_{0j}=-1$ and $\alpha_{1j}=1$.
    \item $w(x, a)=\frac{1}{\pi(x; \pmb{\alpha})}$
\end{itemize}

According to this setting, we have $a_j^{opt}=\frac{-(\hat{\Psi}_{j0}+\hat{\Psi}_{j1} x_j)}{2 \hat{\Psi}_{j2}}$. Note that we consider two forms of model misspecification here. At stage 2, the treatment model was misspecified as the null model while the treatment-free model was correctly specified, whereas at stage 1, the treatment model was correctly specified and the treatment-free model was misspecified by including only $x_1$. Algorithm 2 presents the G-dWOLS code example. According to the obtained results, the estimated blip parameters are $(\widehat{\Psi}_{10}, \widehat{\Psi}_{11}, \widehat{\Psi}_{12}) = (2.5501,  1.0305, -1.1023)$ for stage 1 and $(\widehat{\Psi}_{20}, \widehat{\Psi}_{21}, \widehat{\Psi}_{22}) = (-1.5446,  1.3644, -1.0816)$ for stage 2. Given these estimated blip parameters,   for a patient with covariates $x_1=1$ and $x_2=2$, the estimated DTR prescribes 2.09 units of the drug $(a_1=2.09)$ at stage 1 and $0.55$ units $(a_2=0.55)$ at stage 2.

\end{appendix}

\onecolumn

\begin{tcolorbox}[colback=gray!5!white, colframe=gray!75!black, boxrule=0.6pt, width=1.01\textwidth, arc=4pt, auto outer arc, title=Algorithm 2: G-dWOLS code example (Note: \texttt{gdwols} and \texttt{ipw.fcn} functions are missed)]
\texttt{\text{\textcolor{datacolor}{\#Data Generation}, \textcolor{modelcolor}{Model Specification}, \textcolor{analysiscolor}{DTR estimation}} \\
\textcolor{datacolor}{n <- 10000; k <- 2; m <- 5} \text{\#The total number of individuals, stages, and bins } \\
\textcolor{datacolor}{alpha1 <- (-1, 1), psi1 <- c(1, 1, -1)} \text{\#Stage 1 parameter vectors  $\pmb{\alpha}_1$  and $\pmb{\Psi}_1$}  \\
\textcolor{datacolor}{alpha2 <- (-1, 1), psi2 <- c(1, 1, -1)} \text{\#Stage 2 parameter vectors  $\pmb{\alpha}_2$  and $\pmb{\Psi}_2$}  \\
\textcolor{datacolor}{x1<-abs(rnorm(n,10,1)); x2<-abs(rnorm(n,10,1))}  \text{\#Stages 1 \& 2 covariates} \\
\textcolor{datacolor}{a1<-rnorm(n,alpha1[1]+alpha1[2]*x1,1))} \text{\#Stage 1 treatments as a function of x1}\\
\textcolor{datacolor}{a2<-rnorm(n,alpha2[1]+alpha2[2]*x2,1)} \text{\#Stage 2 treatments as a function of x2} \\
\textcolor{datacolor}{\texttt{gamma1<-as.matrix(cbind(a1,a1*x1,a1\textasciicircum2)) \%*\% psi1}} \text{\#Stage 1 treatment effect}\\
\textcolor{datacolor}{\texttt{gamma2<-as.matrix(cbind(a2,a2*x2,a2\textasciicircum2)) \%*\% psi2}} \text{\#Stage 2 treatment effect}\\
\textcolor{datacolor}{y<-log(x1)+sin(x1)+log(x2)+sin(x2)+gamma1+gamma2 + rnorm(n,0,1)} \text{\#Observed outcome}  \\
\textcolor{datacolor}{data<-data.frame(cbind(y,x1,x2,a1,a2))} \text{\#Data set}  \\
\textcolor{modelcolor}{outcome.mod<-y~1} \text{\# Outcome.mod (correctly specified)}  \\
\textcolor{modelcolor}{blip.mod<-list($\sim$ x1+a1, $\sim$ x2+a2)} \text{\#Blip models (correctly specified)} \\
\textcolor{modelcolor}{treat.mod<-list(a1 $\sim$ x1, a2 $\sim$ 1)} \text{\#Treatment models (Stage 2 misspecified)}  \\
\textcolor{modelcolor}{tf.mod<-list($\sim$ x1, $\sim$ log(x2)+sin(x2))} \text{\#Treatment-free (Stage 1 misspecified)}\\
\textcolor{analysiscolor}{gdwols(outcome.mod, blip.mod, treat.mod, tf.mod, ipw.fcn, data, m, k)}\\
\textcolor{analysiscolor}{\#Blip parameter estimates} \\
\textcolor{analysiscolor}{\#\qquad \qquad psi.hat[[1]]}\\
\textcolor{analysiscolor}{\#   \quad \qquad  2.550130  1.030519 -1.102307   }\\
\textcolor{analysiscolor}{\#\qquad \qquad psi.hat[[2]] }\\
\textcolor{analysiscolor}{\#    \quad \qquad    -1.544591  1.364363 -1.081640   }
}
\end{tcolorbox}  
 
\begin{tcolorbox}[colback=gray!5!white, colframe=gray!75!black, boxrule=0.6pt, width=1.01\textwidth, arc=4pt, auto outer arc, title=Algorithm 3: dWGLM code example (Note: \texttt{dWGLM} function is missed)]
\texttt{\text{\textcolor{datacolor}{\#Data Generation}, \textcolor{modelcolor}{Model Specification}, \textcolor{analysiscolor}{DTR estimation}} \\
\textcolor{datacolor}{expit <- function(x) {1/(1+exp(-x))}} \\
\textcolor{datacolor}{n <- 10000; k <- 2} \text{\#The total number of individuals and stages} \\
\textcolor{datacolor}{psi01 <- -2; psi11 <- 1; psi02 <- -2; psi12 <- 1} \text{\#Stage 1 \& 2  blip  parameters}  \\
\textcolor{datacolor}{x1 <- rnorm(n,2,1); x2 <- rnorm(n,1+ 0.5*x1, 1)}  \text{\#Stages 1 \& 2 covariates} \\
\textcolor{datacolor}{\texttt{a1 <- rbinom(n,1,expit(-5 + x1 + I(x1\textasciicircum2)))}} \text{\#Stage 1 treatments }\\
\textcolor{datacolor}{\texttt{a2 <- rbinom(n,1,expit(-2.5*x2 + sin(x2) + I(x2\textasciicircum2)))}}\text{\#Stage 2 treatments}\\
\textcolor{datacolor}{\texttt{a1opt <- as.numeric(psi01 + psi11*x1 > 0)}} \text{\#Stage 1 optimal decisions}\\
\textcolor{datacolor}{\texttt{reg1 <- (psi01 + psi11*x1)*(a1opt- a1)}} \text{\#Stage 1 regret}\\
\textcolor{datacolor}{\texttt{a2opt <- as.numeric(psi02 + psi12*x2 > 0)}} \text{\#Stage 2 optimal decisions}\\
\textcolor{datacolor}{\texttt{reg2 <- (psi02 + psi12*x2)*(a2opt- a2)}} \text{\#Stage 2 regret}\\
\textcolor{datacolor}{lgtyopt <- x1 + log(abs(x1)) + cos(pi*x1) } \text{\#Observed log of optimal outcome}  \\
\textcolor{datacolor}{p <- expit(lgtyopt - reg1 -reg2)} \text{\#Outcome mean (optimal outcome minus regrets)}  \\
\textcolor{datacolor}{y <- rbinom(n, 1, p)} \text{\#Observed outcome}  \\
\textcolor{datacolor}{mydata <- data.frame(x1, x2,  a1, a2, y)} \text{\#Data set}  \\
\textcolor{modelcolor}{outcome.mod <- y $\sim$ 1} \text{\# Outcome.mod (correctly specified)}  \\
\textcolor{modelcolor}{blip.mod <- list($\sim$ x1, $\sim$ x2)} \text{\#Blip models (correctly specified)} \\
\textcolor{modelcolor}{\texttt{treat.mod <- list(a1$\sim$ x1 + I(x1\textasciicircum2), a2 $\sim$ x2)}} \text{\#Treatment (Stage 1 misspecified)}  \\
\textcolor{modelcolor}{tf.mod <- list($\sim$ x1, $\sim$ x1 * a1 + x2 + I(log(abs(x1))) + I(cos(pi*x1)))} \text{\#Treatment-free models (Stage 2 misspecified)}\\
\textcolor{analysiscolor}{dWGLM(outcome.mod,blip.mod,treat.mod,tf.mod,k,data=mydata)}\\
\textcolor{analysiscolor}{\#\qquad \qquad psi.hat[[1]]}\\
\textcolor{analysiscolor}{\#   \quad \qquad  -0.8373349  0.1856448   }\\
\textcolor{analysiscolor}{\#\qquad \qquad psi.hat[[2]] }\\
\textcolor{analysiscolor}{\#    \quad \qquad   -2.0091954  0.9523452   }
}
\end{tcolorbox}

\begin{tcolorbox}[colback=gray!5!white, colframe=gray!75!black, boxrule=0.6pt, width=1.01\textwidth, arc=4pt, auto outer arc, title=Algorithm 4: DWSurv code example]
\texttt{library("DTRreg") \, \text{\textcolor{datacolor}{\#Data Generation}, \textcolor{modelcolor}{Model Specification}, \textcolor{analysiscolor}{DTR estimation}} \\
\textcolor{datacolor}{set.seed(1)} \\
\textcolor{datacolor}{expit <- function(x) 1/(1 + exp(-x))}  \\
\textcolor{datacolor}{n <- 10000} \text{\#The total number of individuals} \\
\textcolor{datacolor}{x1 <- runif(n, 0.1, 1.29); x2 <- runif(n, 0.9, 2)}  \text{\#Stages 1 \& 2 covariates} \\
\textcolor{datacolor}{a1 <- rbinom(n, 1, expit(2 * x1 - 1))} \text{\#Stage 1 treatments (function of x1)}\\
\textcolor{datacolor}{a2 <- rbinom(n, 1, expit(-2 * x2 + 2.8))} \text{\#Stage 2 treatments (function of x2)} \\
\textcolor{datacolor}{delta <- rbinom(n, 1, expit(-x1 + 2.1))}  \text{\#Censoring indicator  (function of x1)}  \\
\textcolor{datacolor}{beta2 <- c(4, 1.1, -0.2); psi2 <- c(-0.9, 0.6)} \text{\#Stage 2 parameters ($\pmb{\beta}_2$  and $\pmb{\Psi}_2$)}  \\
\textcolor{datacolor}{\texttt{h2beta <- model.matrix($\sim$ x2 + I(x2\textasciicircum3))}} \text{\#Stage 2 non-treatment covariates $\textbf{\textit{h}}_2^{\pmb{\beta}}$}\\
\textcolor{datacolor}{h2psi <- model.matrix($\sim$ x2)} \text{\# Stage 2 tailoring covariates $\textbf{\textit{h}}_2^{\pmb{\Psi}}$}  \\
\textcolor{datacolor}{t2 <- exp(h2beta[delta == 1, ] \%*\% beta2 + a2[delta == 1] * h2psi[delta==1, ] \%*\% psi2 + rnorm(sum(delta), sd = 0.3))} \text{\#Stage 2 survival times vector}  \\
\textcolor{datacolor}{beta1 <-c(6.3, 1.5, -0.8); psi1 <- c(0.1, 0.1)} \text{\#Stage 1 parameters  ($\pmb{\beta}_1$  and $\pmb{\Psi}_1$)}  \\
\textcolor{datacolor}{\texttt{h1beta <- model.matrix($\sim$ x1 + I(x1\textasciicircum4))}} \text{\#Stage 1 non-treatment covariates ($\textbf{\textit{h}}_1^{\pmb{\beta}}$)}\\
\textcolor{datacolor}{h1psi <- model.matrix($\sim$x1)}  \text{\#Stage 1 tailoring covariates matrix $\textbf{\textit{h}}_1^{\pmb{\Psi}}$}  \\
\textcolor{datacolor}{ttilde <- exp(h1beta[delta == 1, ] \%*\% beta1+ a1[delta == 1] *h1psi[delta ==1, ] \%*\% psi1 + rnorm(sum(delta),sd = 0.3))} \text{\#The counterfactual stage 2 survival time under $a_2^{opt}$}  \\
\textcolor{datacolor}{t2opt <- exp(log(t2) + (ifelse(h2psi[delta == 1, ] \%*\%psi2 > 0, 1, 0) -a2[delta == 1]) * h2psi[delta == 1, ]\%*\% psi2)} \text{\#Stage 2 survival time had $a_2^{opt}$ been received}  \\
\textcolor{datacolor}{t1 <- ttilde - t2opt} \text{\# Stage 1 survival time}  \\
\textcolor{datacolor}{c <- rexp(sum(delta == 0), 1/300)} \text{\#Censoring times }  \\
\textcolor{datacolor}{c1 <- runif(sum(delta == 0), 0, c); c2 <- pmax(0, c - c1)} \text{\#Censoring times}  \\
\textcolor{datacolor}{y2 <- y1 <- rep(0, n); y1[delta == 1] <- t1} \text{\#Observed survival times}  \\
\textcolor{datacolor}{y1[delta == 0] <- c1; y2[delta == 1] <- t2; y2[delta == 0] <- c2} \text{\#Observed survival times }  \\
\textcolor{datacolor}{mydata <- data.frame(x1, a1, x2, a2, y1, y2, delta)} \text{\#Data set}  \\
\textcolor{modelcolor}{t <- list($\sim$ y1, $\sim$ y2)} \text{\# Survival models }  \\
\textcolor{modelcolor}{blip <- list($\sim$ x1, $\sim$ x2)} \text{\#Blip models (correctly specified)} \\
\textcolor{modelcolor}{treat <- list(a1 $\sim$ x1, a2 $\sim$ x2)} \text{\#Treatment models (correctly specified)}  \\
\textcolor{modelcolor}{\texttt{tf <-list($\sim$ x1 + I(x1\textasciicircum4),$\sim$ x2 + I(x2\textasciicircum3))}} \text{\#Treatment-free (correctly specified)}\\
\textcolor{modelcolor}{cens <- list(delta $\sim$ x1, delta $\sim$ x1)}  \text{\#Censoring models}\\
\textcolor{analysiscolor}{DWSurv(time = t, blip.mod = blip, tf.mod = tf, treat.mod = treat, cens.mod = cens, var.estim = "bootstrap", data = mydata)}\\
\textcolor{analysiscolor}{\#DTR estimation over  2  stages:}\\
\textcolor{analysiscolor}{\#Blip parameter estimates} \\
\textcolor{analysiscolor}{\#Stage 1 (n = 10000)}\\
\textcolor{analysiscolor}{\#\qquad \qquad Estimate Std\textunderscore Error X95\textunderscore CI\textunderscore Lower X95\textunderscore CI\textunderscore Upper}\\
\textcolor{analysiscolor}{\#a1    \quad \, \,  1.0003    0.0208        0.9595        1.0411 }\\
\textcolor{analysiscolor}{\#x1:a1 \,           0.9904    0.0279        0.9358        1.0451                     }\\
\textcolor{analysiscolor}{\#Stage 2 (n = 10000)}\\
\textcolor{analysiscolor}{\#\qquad \qquad Estimate Std\textunderscore Error X95\textunderscore CI\textunderscore Lower X95\textunderscore CI\textunderscore Upper}\\
\textcolor{analysiscolor}{\#a2    \quad \, \,      0.9578    0.0565        0.8471        1.0686}\\
\textcolor{analysiscolor}{\#x2:a2 \,    1.0351    0.0399        0.9570        1.1133}\\
\textcolor{analysiscolor}{\#Warning: possible non-regularity at stages 1, 2 (probs =  ,  , respectively)}\\
\textcolor{analysiscolor}{\#Recommended dynamic treatment regimen:}\\
\textcolor{analysiscolor}{\#Stage 1: treat if    1.0003  + 0.9904   x1:a1 > 0}\\
\textcolor{analysiscolor}{\#Stage 2: treat if    0.9578  + 1.0351   x2:a2 > 0}
}
\end{tcolorbox}

\twocolumn

\begin{appendix}
\section*{Appendix B: dWGLM code example}\label{appn2} 
 
This section provides an example code for the dWGLM technique presented in Subsection \ref{sec:dWGLM}, where patient covariates are drawn from a normal distribution, and the mean of the second-stage covariate varies with the first-stage covariate.  This setting, along with others considered in the study, is summarized as follows

  \begin{itemize}
  \item Covariates at stages 1 and 2 are $X_1 \sim N(2, 1)$ and $X_2 \sim N(1 + 0.5 X_1, 1)$.
    \item The treatment model at stages 1 and 2 are $\mathbb{P}(A_1=1|x_1)=\left(1+e^{- 5 + x_1 + x_1^2}\right)^{-1}$ and $\mathbb{P}(A_ 2=1|x_ 2)=\left(1+e^{- 2.5 x_2 + x_2^2 + sin(x_2)}\right)^{-1}$, respectively.
    \item The regret model at stage $j=1,2$ is $\mu_j(x_j, a_j; \pmb{\Psi}_j=(-2, 1))= (-2+ x_{j}) (a_j^{opt}-a_j)$.
    \item  The outcome model is also assumed to be
    \begin{eqnarray*}
        logit(\mathbb{P}(Y=1))&=&logit(\mathbb{P}(Y^{opt}=1))\\
        &-&\mu_1(x_1, a_1; \pmb{\Psi}_1)-\mu_2(x_2, a_2; \pmb{\Psi}_2),
    \end{eqnarray*}
    where
    \begin{eqnarray*}
        logit(P(Y^{opt}=1))=x_1+log(|x_1|)+cos(\pi x_1).
    \end{eqnarray*}
\end{itemize}
 
In this example, we introduced a form of model misspecification where the treatment-free model was misspecified at the second stage while the treatment model was correctly specified. Conversely, at the first stage, the treatment model was misspecified, but the treatment-free model was correctly specified. Algorithm  3 provides the dWGLM code example. The estimated blip parameters are $(\widehat{\Psi}_{10}, \widehat{\Psi}_{11}) = (-0.8373,  0.1856)$ for stage 1 and $(\widehat{\Psi}_{20}, \widehat{\Psi}_{21}) = (-2.0092,  0.9523)$ for stage 2. These estimates inform the recommended treatment regimen, which is consistent with the approach used in the dWOLS method.

\end{appendix}

\begin{appendix}
\section*{Appendix C: DWSurv code example}\label{appn3} 
 
We present here the example code for the DWSurv approach reviewed in Subsection \ref{sec:DWSurv}.  For the sake of brevity and without loss of generality, we assume in this scenario that all individuals progress to stage two ($\eta_{2}=1$), with censoring observed exclusively during this stage.  This example assumes all models are correctly specified.  The following settings are used for this example.

\begin{itemize}
    \item The treatment-free model at stages 1 and 2 are $f_1\left(x_1; \pmb{\beta}_1=(6.3, 1.5, -0.8)\right)= 6.3 + 1.5 x_1 - 0.8 x_1^2 $ and $f_2\left(x_2; \pmb{\beta}_2=(4, 1.1, -0.2)\right)= 4 + 1.1 x_2 + x_2^3$, respectively.
    \item The blip model at stages 1 and 2 are $\gamma_1(x_1, a_1; \pmb{\Psi}_1=(0.1, 0.1))=  a_1 (0.1 + 0.1 x_j)$ and $\gamma_2(x_2, a_2; \pmb{\Psi}_2=(-0.9, 0.6))=  a_2 (-0.9 + 0.6 x_2)$, respectively.
    \item The treatment model at stages 1 and 2 are $\mathbb{P}(A_1=1|x_1)=\left(1+e^{- 1 + 2 x_1}\right)^{-1}$ and $\mathbb{P}(A_ 2=1|x_ 2)=\left(1+e^{2.8 - x_2}\right)^{-1}$, respectively.
    \item Censoring model
    \begin{align*}
        g(x, a)=\mathbb{P}(\Delta=1|x)=\left(1+e^{2.1 - x}\right)^{-1}.
    \end{align*}
\end{itemize}

Algorithm 4 shows the DWSurv code example. The blip estimates are displayed by stages along with 95\% confidence intervals shown on the right-hand side for their respective parameters. The estimated blip parameters are $(\widehat{\Psi}_{10}, \widehat{\Psi}_{11}) = (1.0003, 0.9904)$ for stage 1 and $(\widehat{\Psi}_{20}, \widehat{\Psi}_{21}) = (0.9578, 1.0351)$ for stage 2. These estimates guide the recommended treatment regimen summarized at the end of the output.

\end{appendix}

\begin{appendix}
\section*{Appendix D: DWSurvMT code example}\label{appn4} 
 
This section provides the example code for the DWSurvMT approach discussed in \ref{sec:DWSurvMT}. Algorithm 5 shows the DWSurvMT code example.  The data are simulated from an experiment with three available treatment options at each stage, i.e., $N_1 = N_2 = 3$ with two covariates at each stage. We also assume that 80$\%$ of patients enter stage two. The treatment-specified weight is set to be $m_j(a) = N_j^{-1}$, with the inverse probability weight of form $w_j(\mathbf{h}_j, a, 1)=(\pi(\mathbf{h}_j; \pmb{\alpha}) g(\mathbf{h}, a))^{-1}$ for any $a \in \mathcal{A}_j$. It is assumed here that both the treatment-free and weighting models are correctly specified. The censoring time $C$ is generated independently from an exponential distribution $Exp(1/10)$, under which the censoring percentage is about 15\%. For an individual, the baseline covariates $\mathbf{X}_1 =(X_{11}, X_{12})^\top$ follow a normal mixture distribution, while $\mathbf{X}_2 = (1, X_{22})^\top$, the covariates measured prior to the treatment at stage two is an ordinal variable so that $X_{22} \in \{-1, 0, 1\}$ with $\mathbb{P}(X_{22} = -1) = \mathbb{P}(X_{22} = 0) = \mathbb{P}(X_{22} = 1) = 1/3$. The estimated stage-specific targeted blip estimates are displayed at the end of the algorithm.

\end{appendix}

\section*{SUPPLEMENTARY MATERIAL}

The \texttt{R} script files that reproduce the results in Algorithms 1-5 are as follows
\begin{itemize}
\item \texttt{dWOLS.R} : dWOLS code example in Algorithm 1
\item \texttt{G-dWOLS.R} : G-dWOLS code example in Algorithm 2
\item \texttt{dWGLM.R}  : dWGLM code example in Algorithm 3
\item \texttt{DWSurv.R} : DWSurv code example in Algorithm 4
\item \texttt{DWSurvMT.R}  : DWSurvMT code example in Algorithm 5
\end{itemize}

\onecolumn

\begin{tcolorbox}[colback=gray!5!white, colframe=gray!75!black, boxrule=0.6pt, width=1\textwidth, arc=4pt, auto outer arc, title= Algorithm 5: DWSurvMT code example (Note: \texttt{ITRSurv} and \texttt{ipw.fcn}  and \texttt{GenerateData} functions are missed)]
\texttt{library("DTRreg") \, \text{\textcolor{datacolor}{\#Data Generation}, \textcolor{modelcolor}{Model Specification}, \textcolor{analysiscolor}{DTR estimation}} \\
\textcolor{datacolor}{n <- 1000; nx=2; na=3} \text{\#Number of individuals, covariates, and treatments} \\
\textcolor{datacolor}{mydata <- GenerateData1(na,n, gamma)\$mydata} \text{\#Data set} \\
\textcolor{modelcolor}{cen.mod=list(delta $\sim$ x11+x12)} \text{\#Censoring model} \\
\textcolor{modelcolor}{outcome.mod2=logY2 $\sim$ 1} \text{\#Stage 2 outcome model} \\
\textcolor{modelcolor}{blip.mod2=list($\sim$x21+x22)} \text{\#Stage 2 blip model (correctly specified)} \\
\textcolor{modelcolor}{treat.outcome2<-a2 $\sim$ 1} \text{\#Stage 2 treatment outcome model} \\
\textcolor{modelcolor}{\texttt{treat.mod2=list(a2$\sim$x21+x22+I(x22\textasciicircum2))}} \text{\#Stage 2 treatment (correctly specified)}\\
\textcolor{modelcolor}{tf.mod2=list($\sim$ exp(2*x22)+x22)} \text{\#Stage 2 treatment-free (correctly specified)} \\
\textcolor{modelcolor}{outcome.mod1=logY1 $\sim$ 1} \text{\#Stage 1 outcome model} \\
\textcolor{modelcolor}{blip.mod1=list($\sim$ x11+x12)} \text{\#Stage 1 blip model (correctly specified)} \\
\textcolor{modelcolor}{treat.outcome1=a1 $\sim$ 1} \text{\#Stage 1 treatment outcome model} \\
\textcolor{modelcolor}{treat.mod1=list(a1 $\sim$ x11+x12)} \text{\#Stage 1 treatment model (correctly specified)} \\
\textcolor{modelcolor}{\texttt{tf.mod1=list($\sim$ log(abs(x11))+I(x12\textasciicircum2)+x11+x12)}} \text{\#Stage 1 treatment-free model}\\
\textcolor{analysiscolor}{tempout2<-ITRSurv(outcome.mod2,treat.outcome2,blip.mod2,treat.mod2,tf.mod2,\\cen.mod,mydata[mydata\$eta2==1,])}\\
\textcolor{analysiscolor}{out=tempout2\$psi.hat; }\\
\textcolor{analysiscolor}{tempblip=as.matrix(mydata[,c("x21","x22")])\%*\%t(out)}\\
\textcolor{analysiscolor}{tempaopt=apply(tempblip,1,which.max)}\\
\textcolor{analysiscolor}{tempblipopt=tempblip[cbind(seqalong(tempaopt),tempaopt)]}\\
\textcolor{analysiscolor}{currentblip=tempblip[cbind(seqalong(mydata\$a2),mydata\$a2)]}\\
\textcolor{analysiscolor}{logT1pse=log(mydata\$Y1+mydata\$Y2*exp(-currentblip+tempblipopt))}\\
\textcolor{analysiscolor}{mydata[mydata\$eta2==1,"logY1"]=logT1pse[mydata\$eta2==1]}\\
\textcolor{analysiscolor}{tempout1<-ITRSurv(outcome.mod1,treat.outcome1,blip.mod1,treat.mod1,tf.mod1,\\cen.mod,mydata)}\\
\textcolor{analysiscolor}{t(tempout2\$psi.hat)}\\
\textcolor{analysiscolor}{[1,] -0.2848645  0.6622108 -0.3773462}\\
\textcolor{analysiscolor}{[2,] -2.0293590 -0.1926028  2.2219618}\\
\textcolor{analysiscolor}{t(tempout2\$psi.hat)}\\
\textcolor{analysiscolor}{[1,] -0.2848645  0.6622108 -0.3773462}\\
\textcolor{analysiscolor}{[2,] -2.0293590 -0.1926028  2.2219618}\\
}
\end{tcolorbox}

\twocolumn


\newpage

\begin{thebibliography}{4}

\bibitem{bian2023variable}
Bian, Z., Moodie, E. E., Shortreed, S. M., \& Bhatnagar, S. (2023). Variable selection in regression‐based estimation of dynamic treatment regimes. \textit{Biometrics}, 79(2), 988-999.


\bibitem{chakraborty2013statistical}
Chakraborty, B., \& Moodie, E. E. M. \textit{Statistical Methods for Dynamic Treatment Regimes: Reinforcement Learning, Causal Inference, and Personalized Medicine}. Springer, New York, NY, 2013.


\bibitem{chakraborty2014dynamic}
Chakraborty, B., \& Murphy, S. A. (2014). Dynamic treatment regimes. \textit{Annual review of statistics and its application}, 1(1), 447-464.

\bibitem{coulombe2021can}
Coulombe, J., Moodie, E. E., Shortreed, S. M., \& Renoux, C. (2021). Can the risk of severe depression-related outcomes be reduced by tailoring the antidepressant therapy to patient characteristics?. \textit{American Journal of Epidemiology}, 190(7), 1210-1219.


\bibitem{jiang2022doubly}
Jiang, C., Wallace, M., \& Thompson, M. (2022). Doubly-robust dynamic treatment regimen estimation for binary outcomes. \textit{arXiv preprint arXiv:2203.08269}.

\bibitem{jiang2023dynamic}
Jiang, C., Wallace, M. P., \& Thompson, M. E. (2023). Dynamic treatment regimes with interference. \textit{Canadian Journal of Statistics}, 51(2), 469-502.



\bibitem{hong2021individualized}
Hong, Y., Chen, L., Pan, Q., Ge, H., Xing, L., \& Zhang, Z. (2021). Individualized mechanical power-based ventilation strategy for acute respiratory failure formalized by finite mixture modeling and dynamic treatment regimen. \textit{EClinicalMedicine}, 36.



\bibitem{ma2021individualized}
Ma, P., Liu, J., Shen, F., Liao, X., Xiu, M., Zhao, H., ... \& Zhang, Z. (2021). Individualized resuscitation strategy for septic shock formalized by finite mixture modeling and dynamic treatment regimen. \textit{Critical Care}, 25, 1-16.



\bibitem{moodie2014q}
Moodie, E. E., Dean, N., \& Sun, Y. R. (2014). Q-learning: Flexible learning about useful utilities. \textit{Statistics in Biosciences}, 6, 223-243.


\bibitem{moodie2020precision}
 Moodie, E. E., \& Krakow, E. F. (2020). Precision medicine: Statistical methods for estimating adaptive treatment strategies. \textit{Bone Marrow Transplantation}, 55(10), 1890-1896.




\bibitem{moodie2022privacy}
Moodie, E. E., Coulombe, J., Danieli, C., Renoux, C., \& Shortreed, S. M. (2022). Privacy-preserving estimation of an optimal individualized treatment rule: a case study in maximizing time to severe depression-related outcomes.  \textit{Lifetime data analysis}, 28(3), 512-542.




\bibitem{moseley2024personalised}
Moseley, M. J., Stewart, C. E., Bradley, L. C., Fielder, A. R., \& Wallace, M. P. (2024). Personalised versus standardised dosing of occlusion therapy for amblyopia: A randomised controlled trial. \textit{JFO Open Ophthalmology}, 5, 100060.



\bibitem{rich2016optimal}
Rich, B., Moodie, E. E., \& Stephens, D. A. (2016). Optimal individualized dosing strategies: A pharmacologic approach to developing dynamic treatment regimens for continuous‐valued treatments. \textit{Biometrical Journal}, 58(3), 502-517.



\bibitem{robins2004optimal}
Robins, J. M. (2004). Optimal structural nested models for optimal sequential decisions. In \textit{Proceedings of the Second Seattle Symposium in Biostatistics: analysis of correlated data} (pp. 189-326). New York, NY: Springer New York.



\bibitem{simoneau2018non}
Simoneau, G., Moodie, E. E., Platt, R. W., \& Chakraborty, B. (2018). Non-regular inference for dynamic weighted ordinary least squares: understanding the impact of solid food intake in infancy on childhood weight. \textit{Biostatistics}, 19(2), 233-246.



\bibitem{simoneau2020estimating}
Simoneau, G., Moodie, E. E., Nijjar, J. S., Platt, R. W., \& Scottish Early Rheumatoid Arthritis Inception Cohort Investigators. (2020). Estimating optimal dynamic treatment regimes with survival outcomes. \textit{Journal of the American Statistical Association}, 115(531), 1531-1539.



\bibitem{simoneau2020optimal}
Simoneau, G., Moodie, E. E., Wallace, M. P., \& Platt, R. W. (2020). Optimal dynamic treatment regimes with survival endpoints: introducing DWSurv in the R package DTRreg. \textit{Journal of Statistical Computation and Simulation}, 90(16), 2991-3008.



\bibitem{simoneau2020adaptive}
Simoneau, G., Moodie, E. E., Azoulay, L., \& Platt, R. W. (2020). Adaptive treatment strategies with survival outcomes: an application to the treatment of type 2 diabetes using a large observational database. \textit{American journal of epidemiology}, 189(5), 461-469.



\bibitem{schulz2021doubly}
Schulz, J., \& Moodie, E. E. (2021). Doubly robust estimation of optimal dosing strategies. \textit{Journal of the American Statistical Association}, 116(533), 256-268.


\bibitem{spicker2020measurement}
Spicker, D., \& Wallace, M. P. (2020). Measurement error and precision medicine: Error‐prone tailoring covariates in dynamic treatment regimes. \textit{Statistics in Medicine}, 39(26), 3732-3755.

 
\bibitem{wallace2015doubly}
Wallace, M. P., \& Moodie, E. E. (2015). Doubly-robust dynamic treatment regimen estimation via weighted least squares. \textit{Biometrics}, 71(3), 636-644.



\bibitem{wallace2016package}
Wallace, M. P., Moodie, E. E., Stephens, D. A., Simoneau, G., Holloway, S. T., Schulz, J., \& Holloway, M. S. T. (2016). Package ‘DTRreg’. \textit{R package version, 1}.


\bibitem{wallace2017dynamic}
Wallace, M. P., Moodie, E. E., \& Stephens, D. A. (2017a). Dynamic treatment regimen estimation via regression-based techniques: introducing R package DTRreg. \textit{Journal of Statistical Software}, 80, 1-20.


 

\bibitem{wallace2017model}
Wallace, M. P., Moodie, E. E., \& Stephens, D. A. (2017b). Model validation and selection for personalized medicine using dynamic-weighted ordinary least squares. \textit{Statistical Methods in Medical Research}, 26(4), 1641-1653.



\bibitem{zhang2022doubly}
Zhang, Z., Yi, D., \& Fan, Y. (2022). Doubly robust estimation of optimal dynamic treatment regimes with multicategory treatments and survival outcomes. \textit{Statistics in Medicine}, 41(24), 4903-4923.

\bibitem{zhang2020individualized}
Zhang, Z., Zheng, B., \& Liu, N. (2020). Individualized fluid administration for critically ill patients with sepsis with an interpretable dynamic treatment regimen model.  \textit{Scientific Reports}, 10(1), 17874.

\end{thebibliography}
\end{document}